\documentclass[preprint,prd,showpacs]{revtex4}

\usepackage{amsmath}
\usepackage{amsfonts}
\usepackage{amssymb}
\usepackage{bbm}
\usepackage{latexsym}
\usepackage[dvips]{graphicx}
\usepackage{graphics}
\usepackage{color}
\definecolor{watpliwy}{rgb}{1,0,0}

\begin{document}

\title{Stability of FRW Cosmology with Generalized Chaplygin Gas} 
\author{Marek Szyd{\l}owski}
\email{uoszydlo@cyf-kr.edu.pl}
\affiliation{Astronomical Observatory, Jagiellonian University,\\ Orla 171, 30-244 Krak\'ow, Poland}
\author{Wojciech Czaja}
\email{czaja@oa.uj.edu.pl}
\affiliation{Astronomical Observatory, Jagiellonian University,\\ Orla 171, 30-244 Krak\'ow, Poland}
\date{\today}

\begin{abstract}
We apply methods of dynamical systems to study the behaviour of universe dominated by the generalized
Chaplygin gas. We reduce the dynamics to a 2--dimensional Hamiltonian system and study its behaviour for 
various ranges of parameters. The dynamics is studied on the phase plane by using methods of qualitative 
analysis of differential equations. The behaviour of trajectories at infinity is studied in some convenient 
coordinates introduced on the phase plane. Hence we shown that FRW model with the generalized 
Chaplygin gas is structurally stable. We clearly find the domains of cosmic acceleration as well as 
conditions for which the horizon problem is solved. We also define some general class of fluids which 
generalize the Chaplygin gas. The dynamics of such models in terms of energy conditions is also discussed.
\end{abstract}


\maketitle

\section{Introduction}
The generalized Chaplygin gas \cite{kamenshchik} was recently proposed as a alternative to the cosmological
constant in explaining the accelerating universe. The equation of state, describing the 
pressure $p$ of this gas, is given in terms of the energy density $\rho$ by the relation 
$p=-A/\rho^{\alpha}$, where $A$ is a positive constant and $0 < \alpha \leqslant 1$. An universe filled by 
matter in the form of Chaplygin gas evolves from the phase dominated by dust to a de Sitter phase where 
$p \simeq -\rho$, $\rho = \Lambda$ through an intermediate regime described by the equation of state for 
the Zeldovich stiff matter with $p=\rho$.

There are many various ways in which one can explain the transition from a matter dominated universe to one
with the accelerated expansion, for example brane inspired models \cite{avelino} or vacuum metamorphosis
\cite{bassett}. The generalized Chaplygin gas is a generalization of Chaplygin gas \cite{bilic} motivated
by considering a d-brane in a spacetime of d+2 dimensions. We should note that there is no such motivation
for the generalized Chaplygin gas. At a toy model, one can simply generalize equation of state for the 
Chaplygin gas $p=-A/\rho$ to the more general dependence on density $p=-A/\rho^{\alpha}$ \cite{bento}.
Interesting future of this model is that it is naturally provided explanation of transition from a 
decelerating phase at early stage of evolution to accelerated expansion at later stages.

There are monting evidences that our universe accelerates at present \cite{perlmutter}. Therefore it should 
be dominated by some kind of matter with negative pressure, the so--called dark energy. The different dark 
energy models are motivated by the recent measurements of high--redshift supernovae Ia observations, which 
cannot be explained by the canonical E-deS model. The simplest model of dark energy is the cosmological 
constant but in this case we cannot explain why we don't observe the large energy density 
$\rho_{\Lambda} \equiv \frac{\Lambda}{8 \pi G} \simeq 10^{76} {\rm GeV}$, expected from particle physics
which is about $10^{123}$ times larger than the value predicted by the FRW model.

Therefore, while the most obvious candidate for such energy seems to be vacuum energy, we are looking for
alternatives which satisfies the equation of state for the quintessential matter $p = w(z)\rho$ for 
simplicity, where $w(z) \equiv p/\rho$ is a quintessential parameter as a function of redshift $z$. In this 
context we consider the universe filled by the generalized Chaplygin gas as a possible candidate for dark 
energy.

In investigation of full dynamics we use the dynamical systems methods to analyse the global properties of 
the system. We visualize all evolutional paths for all possible initial conditions as curves (and singular
solutions) in the phase plane $(H,\rho)$, where $H = \frac{d}{dt}(\ln{a})$ is the Hubble function. The phase
space diagrams are very useful both to analyze the assymptotic states, their stability and to study other 
interesting properties of time evolution of models under studies. Especially, it is interesting to study the 
solution of horizon problem and initial conditions for acceleration of our universe.

As complementary to phase space analysis, the Hamiltonian approach is developed. In this approach the 
dynamics is reduced to the 1-dimensional Hamiltonian system. Hence, all possible evolutional paths are 
classified in the configurational space $\{a \colon a \geqslant 0\}$. The construction of the Hamiltonian is 
very advantageous and can be applied not only for quantum cosmology. It is possible to make the qualitative 
classification of evolutional paths by analyzing the characteristic curve which represents the boundary of 
the domain admissible for motion of the system.

The next advantage of representing dynamics in terms of Hamiltonian is possibility to discuss the 
stability of critical points which is based only on the geometry of the potential function.

The presented formalism gives us a natural base to discuss the property of structural stability of the model.
We give a simple proof of structural stability of the FRW model with the generalized Chaplygin gas.

The structural stability is sometimes considered as a precondition of the ``physical existence". 
The existence of many drastically different mathematical models well agreeing with observational data (taking
into account final measurement errors) seems to be fatal for the empirical methods of modern physics 
\cite{golda}. Therefore, any structurally unstable models are probably not physically meaningful (such as 
some of discussed in the present work). This suggests -- if we agree that physically realistic may be only 
what is structurally stable -- that the Chaplygin gas (or other component type of possitive cosmological 
constant), in the real universe, should be different from zero.
In this paper we also demonstrate the effectiveness of analysis of dynamics in terms of one--dimensional
Hamiltonian flow to answer the question: How are trajectories with interesting properties distributed
in the phase plane? Along which trajectories is the acceleration condition, 
$\ddot{a} = -\partial V/\partial a > 0$ satisfied? Are the trajectories, along which cosmological problems
are solved, distributed in some typical or exceptional way? It is the problem of degree of generality for 
interesting properties.

Organization of the text is following. In section 2 the FRW model with the generalized Chaplygin gas is 
investigated in terms of dynamical systems methods. In section 3 we find that the models with Chaplygin gas 
are structurally stable. In section 4 we consider the dynamics of FRW model with the generalized Chaplygin 
gas as a 1-dimensional Hamiltonian system and prove the structural stability of this model. Section 4 gives 
a summary of results and comments on them. In section 5 we discuss how the model with the generalized 
Chaplygin gas fits the SNIa data. 

\section{The FRW model with the Chaplygin gas as a two-dimensional dynamical system}

The main aim of qualitative analysis of differential equation is investigation of global dynamics in the 
phase plane for all possible initial conditions instead finding its explicit solutions. The solution of 
dynamical system $\dot{x}=f(x)$, $f \in C^{\infty}$ is given by a function of two variables, an initial 
condition $x_{0}$ as well as the time $t$ $\rightarrow$ $x(x_{0},t)$. The space of all states $x$ of the 
system at given time is called phase space. We reduce dynamics to the 2-dimensional phase space in which 
singular solutions $\dot{x}=0$ are represented by critical points and nonsingular ones by phase curves. 
In this representation the phase diagrams in a two-dimensional phase space allow to analyze the acceleration 
and the horizon problem in a clear way. We reduce the dynamics to the two-dimensional phase space with an 
autonomous sytem of equations $\dot{x}=P(x,y)$, $\dot{y}=Q(x,y)$, where $x$, $y$ are coordinates,
$P$, $Q \in C^{\infty}$, and a dot denotes the differentiation with respect to cosmological time $t$.

The classification of non-degenerate critical points can be given in terms of eigenvalues of a linearization 
matrix at a critical point. The eigenvalues $\lambda_{1}$, $\lambda_{2}$ are invariants of critical points,
i.e., they do not change as we change the coordinates $x$, $y$. We call a critical point $(x_{0},y_{0})$
non-degenerated (hyperbolic), if ${\rm Re}\lambda_{1}~\neq~0$ and ${\rm Re}\lambda_{2}~\neq~0$. 
The particular trajectories of the system approach a critical point for $t \rightarrow \infty$ or escape away
from it for $t \rightarrow -\infty$ along a direction vector ${\bf k}=(k_{x},k_{y})$. These directions are 
simply eigenvectors at the critical point $(x_{0},y_{0})$. If $\lambda_{1} \neq \lambda_{2}$, then a slope
of a tangent vector to trajectories as they are reaching this critical point is $k = k_{x}/k_{y}$ given
in terms of eigenvalues
\begin{equation}
k_{1} = \frac{\lambda_{1} - P'_{x}(x_{0},y_{0})}{P'_{y}(x_{0},y_{0})}, \hspace{5mm}
k_{2} = \frac{\lambda_{2} - P'_{x}(x_{0},y_{0})}{P'_{y}(x_{0},y_{0})}
\label{keigen}
\end{equation}
and a prime denotes the differentiaion with respect to $x$ or $y$.

The behaviour of the system in the neighbourhood of the critical point $(x_{0},y_{0})$ is qualitatively 
equivalent to the behaviour of its linear part
\begin{align}
\label{linp1}
\dot{x}&=P'_{x}(x_{0},y_{0})(x-x_{0})+P'_{y}(x_{0},y_{0})(y-y_{0}) \\
\label{linp2}
\dot{y}&=Q'_{x}(x_{0},y_{0})(x-x_{0})+Q'_{y}(x_{0},y_{0})(y-y_{0}).
\end{align}

After integration, system (\ref{linp1})-(\ref{linp2}) gives
\begin{align}
\label{linpr1}
x-x_{0}&={\rm Re}(C_{1}e^{\lambda_{1}t}+C_{2}e^{\lambda_{2}t})\\
\label{linpr2}
y-y_{0}&={\rm Re}(C_{1}k_{1}e^{\lambda_{1}t}+C_{2}k_{2}e^{\lambda_{2}t}).
\end{align}
These formulas (\ref{keigen})-(\ref{linpr2}) allow to study the behaviour of the system near its 
non-degenerate critical points and then to draw global phase portraits of a dynamical system. 
For completeness of analysis the behaviour at infinity is considered. For this aim it is useful to compactify
$\mathbb{R}^{2}$ by adding a circle at infinity. One can do that simply by introducing the projective 
coordinates.

The two maps cover the circle at infinity $x~=~\infty$, $y~=~\infty$:
\begin{align}
\label{infmap1}
z&=\frac{1}{x},& u&=\frac{y}{x};& z&=0,& -\infty<u<+\infty \\
\intertext{and}
\label{infmap2}
v&=\frac{1}{y},& w&=\frac{x}{y};& v&=0,& -\infty<w<+\infty. 
\end{align}

Let us consider now the FRW model with the vanishing $\Lambda$-term in the following representation

\begin{subequations}
\begin{align}
\dot{H}& \equiv \frac{dH}{dt}=-H^{2}-\frac{1}{6}(\rho+3p)=P(H,\rho),\label{difsys1} \\
\dot{\rho}& \equiv \frac{d\rho}{dt}=-3H(\rho+p)=Q(H,\rho),\label{difsys2} 
\end{align}
\label{difsys}
\end{subequations}
where all symbols have their usual meanings. The first integral of (\ref{difsys1}) is $\rho=3H^{2}+3k/a^{2}$,
where $k$ is curvature and $a$ is a scale factor.

The equation of state for the generalized Chaplygin gas
\begin{align}
p&= -\frac{A}{\rho^{\alpha}},\hspace{5mm} A > 0,\hspace{3mm} 0 < \alpha \leqslant 1
\label{chapstate}
\end{align}
has been assumed. In equation (\ref{difsys1}), $(H,\rho)$ are chosen as phase variables. Because the 
dependence $p$ on $H$ signifies, the bulk viscosity in general $p=p(H,\rho)$ can be postulated. The evolution
of the system is represented by trajectories in the $(H,\rho)$-space uniquely determined by the initial 
conditions $(H_{0},\rho_{0})$.

The first equation from the system (\ref{difsys1}) can be recognized as the Raychaudhuri equation for the 
congruence of the world lines of matter in spacetime with the R-W symmetry. The second one is a condition of 
the conservation energy-momentum tensor.

Let us introduce the following sets (SEC) \& (WEC)
\begin{subequations}
\label{sekwek}
\begin{align}
W &= \{(H,\rho)\colon \rho+p \geqslant 0 \},\label{sekweka}\\
S &= \{(H,\rho)\colon \rho+3p \geqslant 0 \}\label{sekwekb}.
\end{align}
\end{subequations}

System (\ref{difsys1}) has the critical points of two types:
\begin{enumerate}
\item
static $(H=0)$ critical points of the system (\ref{difsys}) are determined by intersection of the 
$\rho$-axis and the boundary $\partial S$ of the $S$ set, i.e. 
$\{{\rm static\  critical\  points}\}=\{\rho\ {\rm axis}\} \cap \partial S $; 
\item
nonstatic critical points of (\ref{difsys}) are situated on the intersection of the flat FRW trajectory
$\rho = 3H^{2}$ and the boundary of $W$, i.e.
$\{{\rm nonstatic\  critical\  points}\}=\{{\rm trajectory}\ k=0\} \cap \partial W$.
\end{enumerate}

From the physical point of view critical points on the phase space corresponds to an asymptotic states
of the universe. To determine its character (i.e. a type of stability) we consider linearization of 
(\ref{difsys}) at the critical point $(H_{0},\rho_{0})$. The linearization matrix of the system is

\begin{equation}
\mathcal{A} = \left[
\begin{matrix}
\frac{\partial P}{\partial H},& \frac{\partial P}{\partial \rho}\\
\frac{\partial Q}{\partial H},& \frac{\partial Q}{\partial \rho}\\
\end{matrix}
\right]_{(H_{0},\rho_{0})} 
=\left[
\begin{matrix}
-2H,& -\frac{1}{6}\frac{d}{d \rho}(\rho + 3p)\\
-3(\rho + p),& -3H\frac{d}{d \rho}(\rho + p)\\
\end{matrix}
\right]_{(H_{0},\rho_{0})}
\label{linmat1}
\end{equation}

The characteristic equation ${\rm det}(\mathcal{A}-\lambda {\mathbbm 1})=0$ takes the form
\begin{equation}
\lambda^{2}-({\rm Tr}\mathcal{A})\lambda+{\rm det}\mathcal{A}=0,
\label{chareq}
\end{equation}
where
\begin{equation}
{\rm Tr}\mathcal{A} = 0,\hspace{3mm} {\rm det}\mathcal{A}=-\frac{1}{2}(\rho+p)_{0}\frac{d}{d\rho}\Bigg|_{0}(\rho+3p)
\label{critpcond1}
\end{equation}
for static critical point and
\begin{equation}
{\rm Tr}\mathcal{A} = -H_{0}\Bigg[2+3\frac{d}{d\rho}\Bigg|_{0}(\rho+p)\Bigg],\hspace{3mm} {\rm det}\mathcal{A} = 6H_{0}^{2}\frac{d}{d\rho}\Bigg|_{0}(\rho+p)
\label{critpcond2}
\end{equation}
for non-static critical points; here $0$ denotes the value of the expresion at the critical point 
$(H_{0},\rho_{0})$.

Let us assume now that 1) the phase space does not contain trajectories along which the energy condition 
$\rho + p \geqslant 0$ is violated in a finite value of $t$. In terms of dynamical system this assumption
is equivalent to the condition that $\partial W$ is a trajectory in the phase plane since $\dot{\rho} = 0$
on $\partial W$. We also postulate that 2) $\rho + p \geqslant 0$ only for $\rho > \rho_{\rm min}$. On the 
strength of the above assumptions the evolution of the universe is described by one of the trajectories lying
between $\rho_{\rm min}$ and infinity and no trajectory intersects $\partial W$ in a finite value of time.
Every model violates the strong energy condition.

In this case there is only one static critical point of the saddle type (then eigenvalues are real and 
opposite signs) and two non-static critical points, stable node if $H_{0}>0$ and unstable node if $H_{0}<0$. 
They represents deS types of evolution.

\begin{figure}[ht]
\begin{center}
$\begin{array}{c@{\hspace{0.2in}}c}
\multicolumn{1}{l}{\mbox{\bf a)}} & 
\multicolumn{1}{l}{\mbox{\bf b)}} \\ [-0.5cm]
\includegraphics[scale=0.7, angle=270]{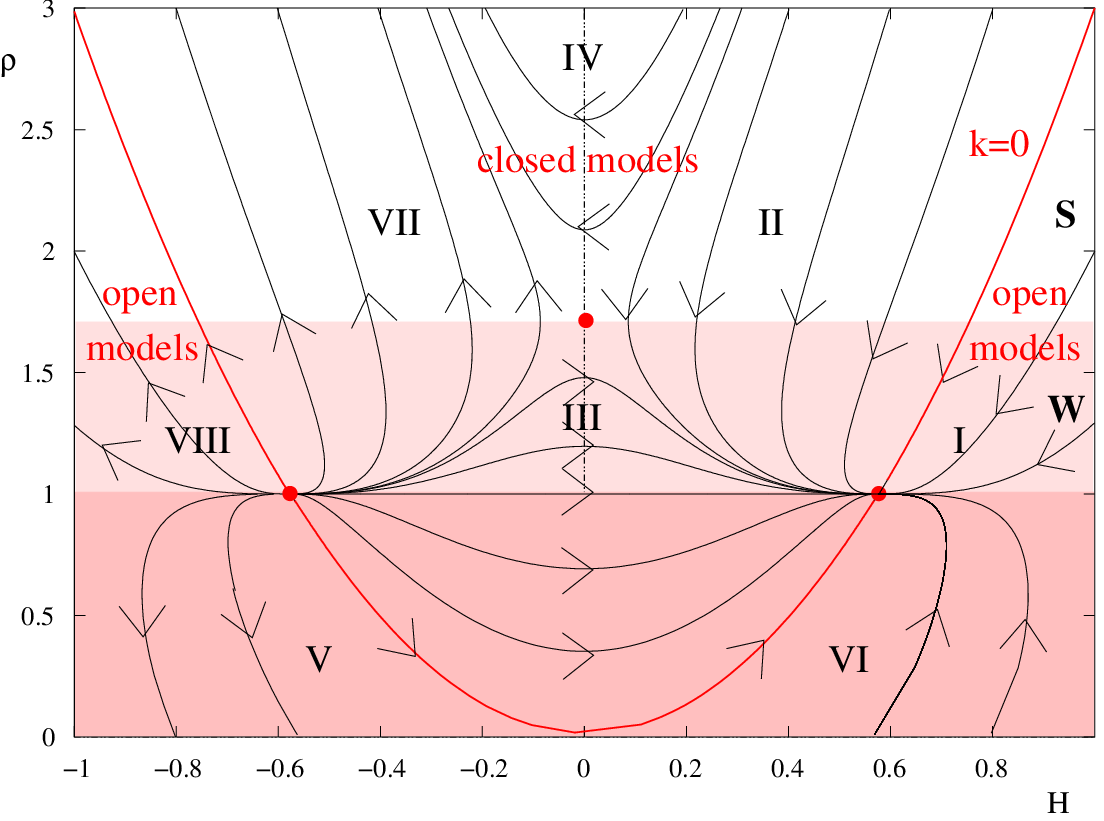} & 
\includegraphics[scale=0.7, angle=270]{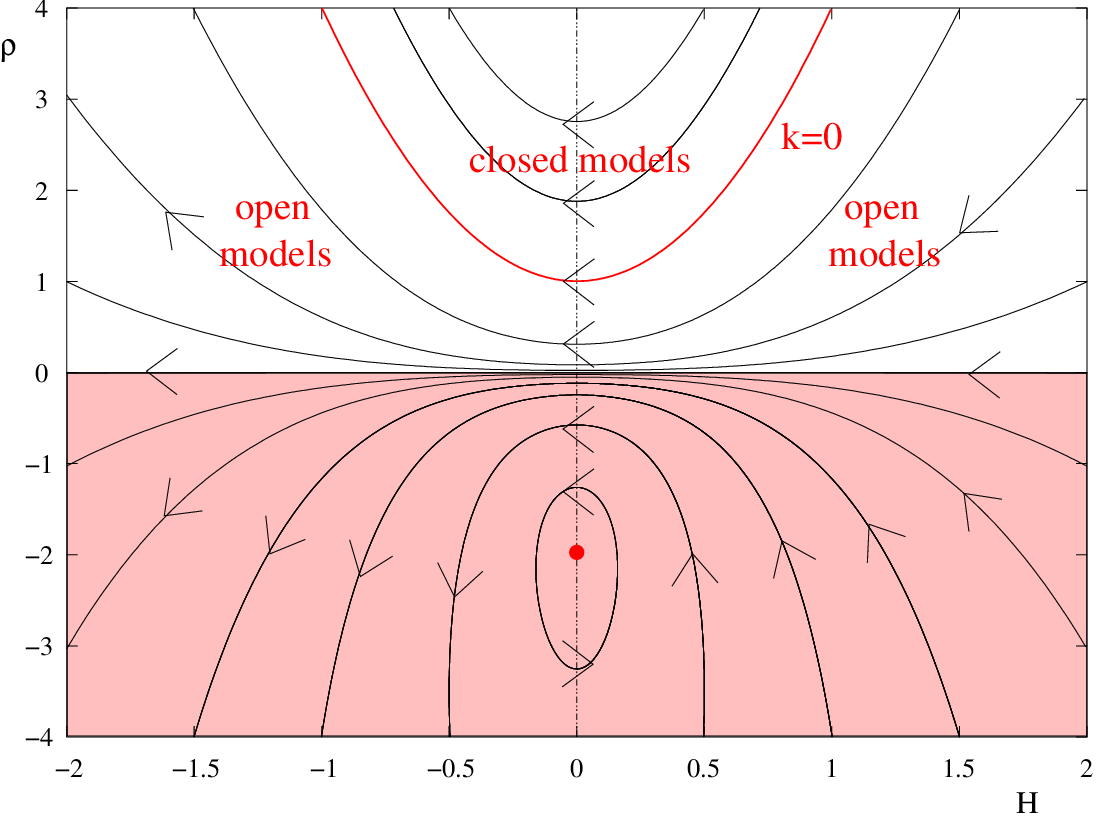} \\ [0.4cm]
\multicolumn{1}{l}{\mbox{\bf c)}} & 
\multicolumn{1}{l}{\mbox{\bf d)}} \\ [-0.5cm]
\includegraphics[scale=0.7, angle=270]{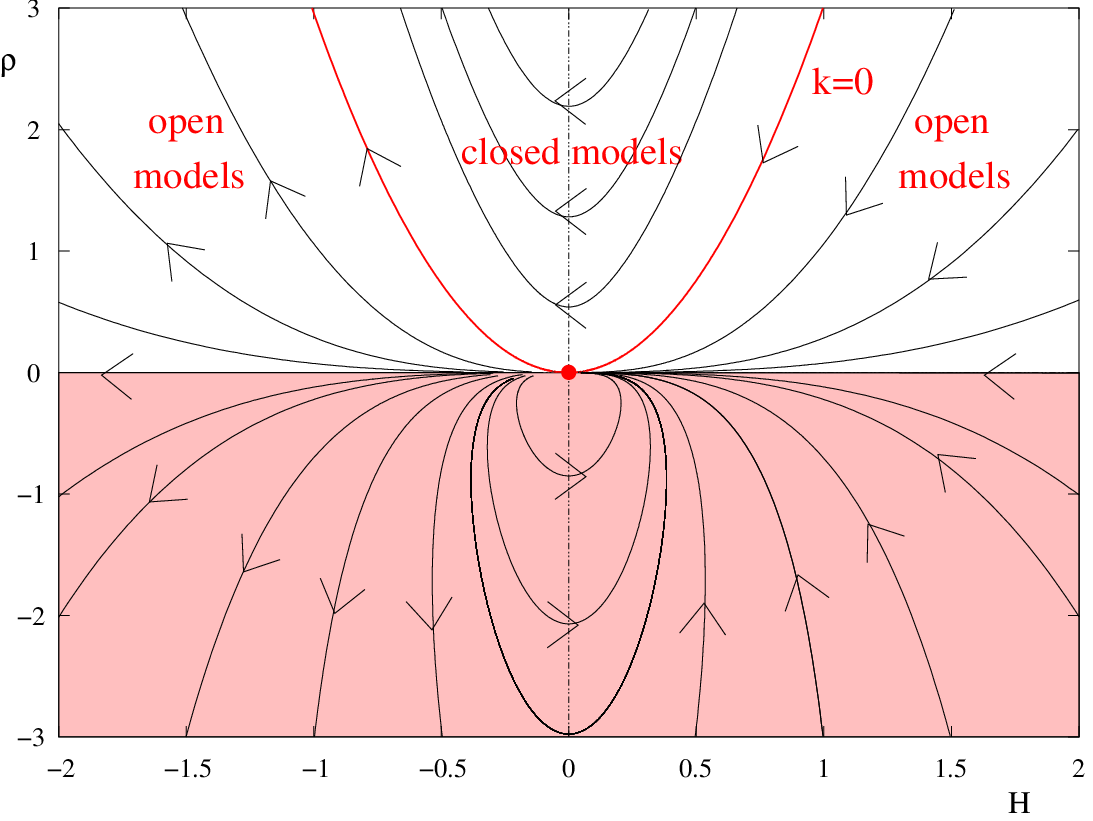} & 
\includegraphics[scale=0.7, angle=270]{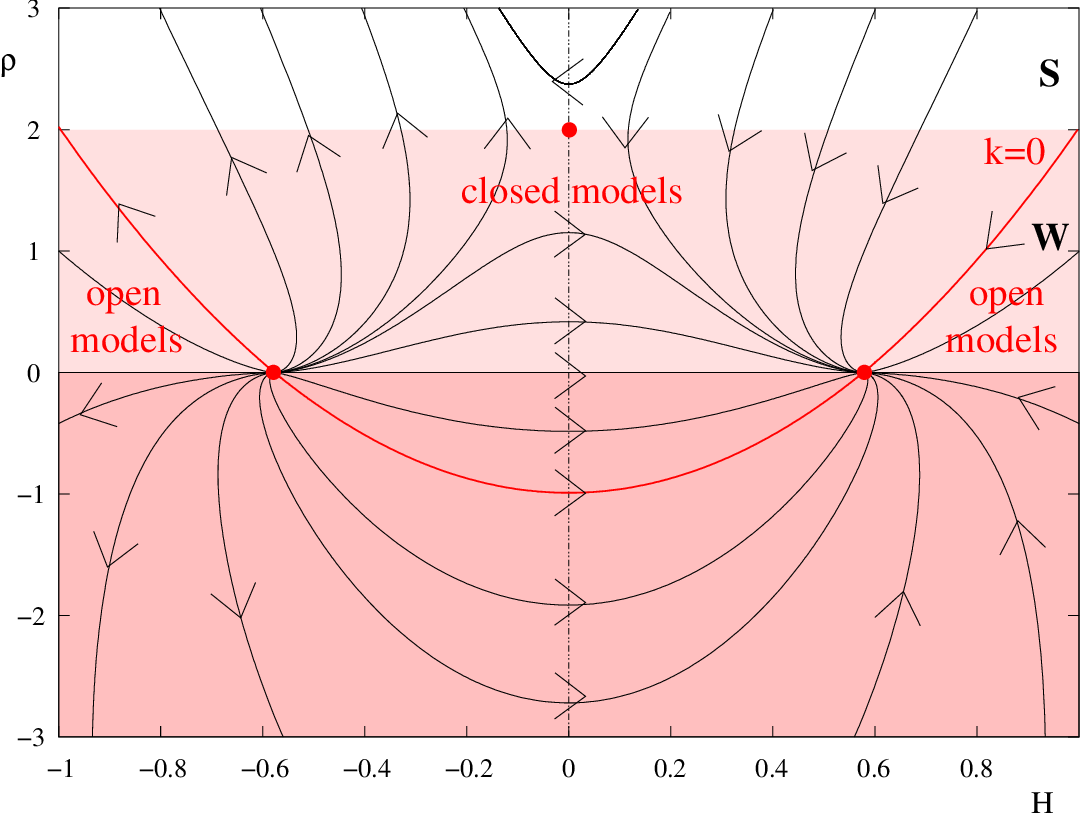} \\ [0.4cm]
\end{array}$
\end{center}
\caption{The phase portraits for {\bf a)} the FRW model with Chaplygin gas ($A=1$,$\alpha=1$) 
and the dust filled FRW models with cosmological constant {\bf b)} $\Lambda<0$, {\bf c)} $\Lambda=0$ and 
{\bf d)} $\Lambda>0$. Let us note the equivalence of phase portraits from figures {\bf a)} and {\bf d)}
in the domains $\rho \geqslant A^{1/(1+\alpha)}$ and $\rho \geqslant 0$ respectively.
}
\label{figababbbc}
\end{figure}

The phase diagram of such a model is shown in Fig.~\ref{figababbbc}a. For comparison the phase portraits for 
dust filled models with the cosmological constant are also presented (Fig.~\ref{figababbbc}b-d). 
In these cases
$$ \dot{H} = -H^{2} - \frac{1}{6}\rho + \frac{\Lambda}{3},\hspace{3mm} \dot{\rho} = -3H\rho. $$

Let us note that all flat and open solutions experience acceleration for the late time after crossing
line $\{\rho = \rho_{\rm crit}\}$ which corresponds to the energy density for the Einstein static universe.
All models which are situated close to the flat model also accelerate.

There are some characteristic domains in Fig.~\ref{figababbbc}a but both the model with generalized Chaplygin
gas (Fig.~\ref{figababbbc}a) and the model with positive cosmological constant $\Lambda$ 
(Fig.~\ref{figababbbc}d) have equivalent portraits in their physical domains: 

\noindent
I -- open models are starting from a singularity and going towards the stable deS node\\
II -- closed models are starting from a singularity and going to the global attractor -- the stable deS node\\
III -- trajectories moving in this region correspond to the closed models starting from the unstable deS
node towards the stable deS node after bounce.\\
IV -- the trajectories situated in region IV are confined by the separatrix going from the singularity
to the saddle point and by the separatrix coming out from the saddle point. This region is covered by 
trajectories describing the closed models evolving from the initial singularity to the maximal size
and recolapsing to the singularity.\\
V -- the trajectories of region V decscribe the open models which begin their evolution from the unstable
deS node and end on line $\rho = 0$ on which the transformation of time is singular.\\
VI -- in this region trajectories go out of the stage $\rho = 0$ to the stable node.\\

Let us note that because the phase portrait is symmetric when $H$ changes its sign, in the regions V-VI,
I-VIII and II-VII the picture is quite symmetric.

We consider our system (\ref{difsys}) in the physical region defined as 
\begin{equation*}
\{(H,\rho)\colon \rho \geqslant A^{1/(1+\alpha)}; A > 0, 0 < \alpha \leqslant 1\}.
\end{equation*}

For the system under consideration the relation $\rho(a)$ can be obtained from (\ref{difsys2}).
Integration gives
\begin{equation}
\rho(a) = \Bigg(A+\frac{B}{a^{3(1+\alpha)}}\Bigg)^{\frac{1}{1+\alpha}}.
\label{chapqa}
\end{equation}

Relation (\ref{chapqa}) interpolates between a universe dominated by dust and de Sitter one via a phase
described by the soft matter equation of state $p = \gamma \rho$. It would be useful to 
rewrite (\ref{chapqa}) in the form
\begin{subequations}
\begin{align}
\rho(a)&=\rho_{0,{\rm Chapl}}\Bigg(A_{s}+\frac{1-A_{s}}{a^{3(1+\alpha)}}\Bigg)^{\frac{1}{1+\alpha}},
\label{chapsta} \\
\intertext{or}
\rho(a)&=\rho_{0,{\rm Chapl}}\Big(A_{s}+(1-A_{s})(1+z)^{3(1+\alpha)}\Big)^{\frac{1}{1+\alpha}},
\label{chapstb} 
\end{align}
\end{subequations}
where $1+z = a^{-1}$, $A_{s}=A/(\rho_{0,{\rm Chapl}})^{1+\alpha}$ and 
$\rho_{0,{\rm Chapl}}=(A+B)^{1/(1+\alpha)}$.

Let us notice that even though equation (\ref{chapstate}) admits a wider range of positive $\alpha$ the
sound velocity $c_{s}^{2} \equiv dp/d\rho = \alpha A/\rho^{1+\alpha}$ does not exceed the velocity of light. 
Therefore we assume $0< \alpha \leqslant 1$, $B>0$ and the case of $\alpha=1$ corresponds to the Chaplygin 
gas.

The first integral of (\ref{difsys}) takes the form
\begin{align}
3H^{2}&=\rho-\frac{3k}{a^{2}}=
\rho_{0,{\rm Chapl}}\Bigg(A_{s}+\frac{1-A_{s}}{a^{3(1+\alpha)}}\Bigg)^{\frac{1}{1+\alpha}}-\frac{3k}{a^{2}},
\label{fintchap1} \\
\intertext{or}
H^{2}&=H_{0}^{2}\Big\{\Omega_{{\rm Chapl},0}\Big(A_{s}+(1-A_{s})(1+z)^{3(1+\alpha)}\Big)^{\frac{1}{1+\alpha}}
+\Omega_{k,0}(1+z)^{2}\Big\},
\label{fintchap2} 
\end{align}
where $\Omega_{{\rm Chapl},0}=\rho_{0,{\rm Chapl}}/3H_{0}^{2}$ is the present density parameter
for the generalized Chaplygin gas. If we substitute into (\ref{fintchap2}) $z = 0$ ($H = H_{0}$) then we 
obtain the constraint $\Omega_{{\rm Chapl},0} + \Omega_{k,0} = 1$.

Let us note that trajectories of the FRW model with generalized Chaplygin gas lie on the algebraic curves
given by the relation
\begin{equation}
3k\Bigg(\frac{\rho^{1+\alpha}-A}{B}\Bigg)^{\frac{2}{3(1+\alpha)}} = \rho-3H^{2},
\label{chapac}
\end{equation}
where $\rho>3H^{2}$ for $k=+1$, $\rho=3H^{2}$ for $k=0$ and $\rho<3H^{2}$ for $k=-1$; 
where $\rho \geqslant A^{\frac{1}{1+\alpha}}$.

Therefore the phase space is divided by the parabolic curve $\rho = 3H^{2}$ on different domains with respect
to the curvature index. The flat model trajectory $\{k = 0\}$ separates the regions of the model with the 
negative and positive curvature.

After the substitution of the form of equation of state (\ref{chapstate}) into (\ref{difsys}) and the 
reparametrization of time $t$ such that a new time variable is a monotonic  function of original cosmological
time we obtain the equivalent dynamical system
\begin{subequations}
\begin{align}
\frac{dH}{d\tau}&=-H^{2}\rho^{\alpha}-\frac{1}{6}(\rho^{\alpha+1}-3A)=P(H,\rho), \label{newdifsys1} \\
\frac{d\rho}{d\tau}&=-3H(\rho^{\alpha+1}-A)=Q(H,\rho), \label{newdifsys2}
\end{align}
\label{newdifsys}
\end{subequations}
where $t \rightarrow \tau \colon dt/\rho^{\alpha}=d\tau$. 
Of course relation (\ref{chapac}) plays role of the first integral for (\ref{newdifsys}).

System (\ref{newdifsys}) is a special case of the systems for which the weak energy condition is satisfied
only for $\rho > \rho_{\rm min} = A^{1/(1+\alpha)}$. Therefore we have critical points of two types
\begin{enumerate}
\item
static $H_{0}=0$, $\rho_{0}=(3A)^{\frac{1}{1+\alpha}}$
\item
nonstatic $H_{0}=\frac{\sqrt{3}}{3} A^{\frac{1}{2(1+\alpha)}}$, $\rho_{0}=A^{\frac{1}{1+\alpha}}$
\end{enumerate}
The critical points lie in the physical region 
$\{(H,\rho)\colon \rho \geqslant A^{\frac{1}{1+\alpha}}$, $H \in {\mathbb R}\}$.
From the physical point of view the static critical point is representing the static Einstein universe 
whereas non-static one is deS solution. For the static critical point ${\rm Tr}A=0$ and 
${\rm det}A=-\frac{A(1+\alpha)}{\rho_{0}^{\alpha}}$. Therefore the characteristic equation for eigenvalues
of the linearization matrix admits for this case only real solutions of opposite signs. This means that 
static critical point is a saddle point for any value of $\alpha$ and a positive value of constant $A$. For 
the non-static critical point ($H_{0} \neq 0$) we have:
\begin{gather*}
{\rm Tr}A = -H_{0}(5+3\alpha),\\
{\rm det}A = 6H_{0}^{2}(1+\alpha),\\
\Delta = ({\rm Tr}A)^{2} - 4{\rm det}A = H_{0}^{2}(9\alpha^{2}+6\alpha+1)>0.
\end{gather*}
Therefore for any $\alpha$ we have stable ($H_{0}>0$) and unstable ($H_{0}<0$) nodes. The structure of phase 
space does not depend on the particular choice of $\alpha \in (0,1]$.
\begin{figure}[ht]
\begin{center}
$\begin{array}{c@{\hspace{0.2in}}c}
\multicolumn{1}{l}{\mbox{\bf a)}} & 
\multicolumn{1}{l}{\mbox{\bf b)}} \\ [-0.5cm]
\includegraphics[scale=0.5]{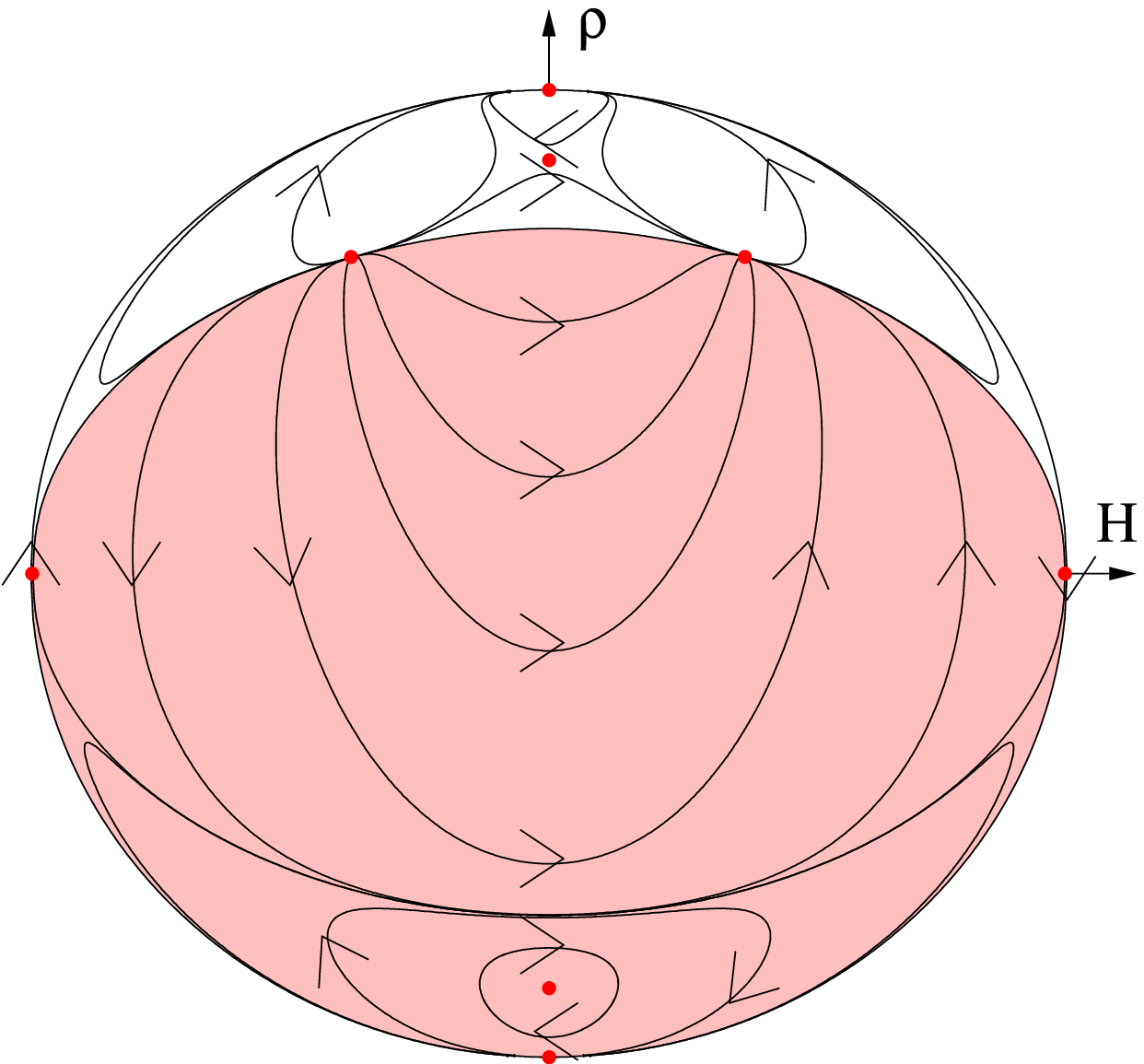} & 
\includegraphics[scale=0.5]{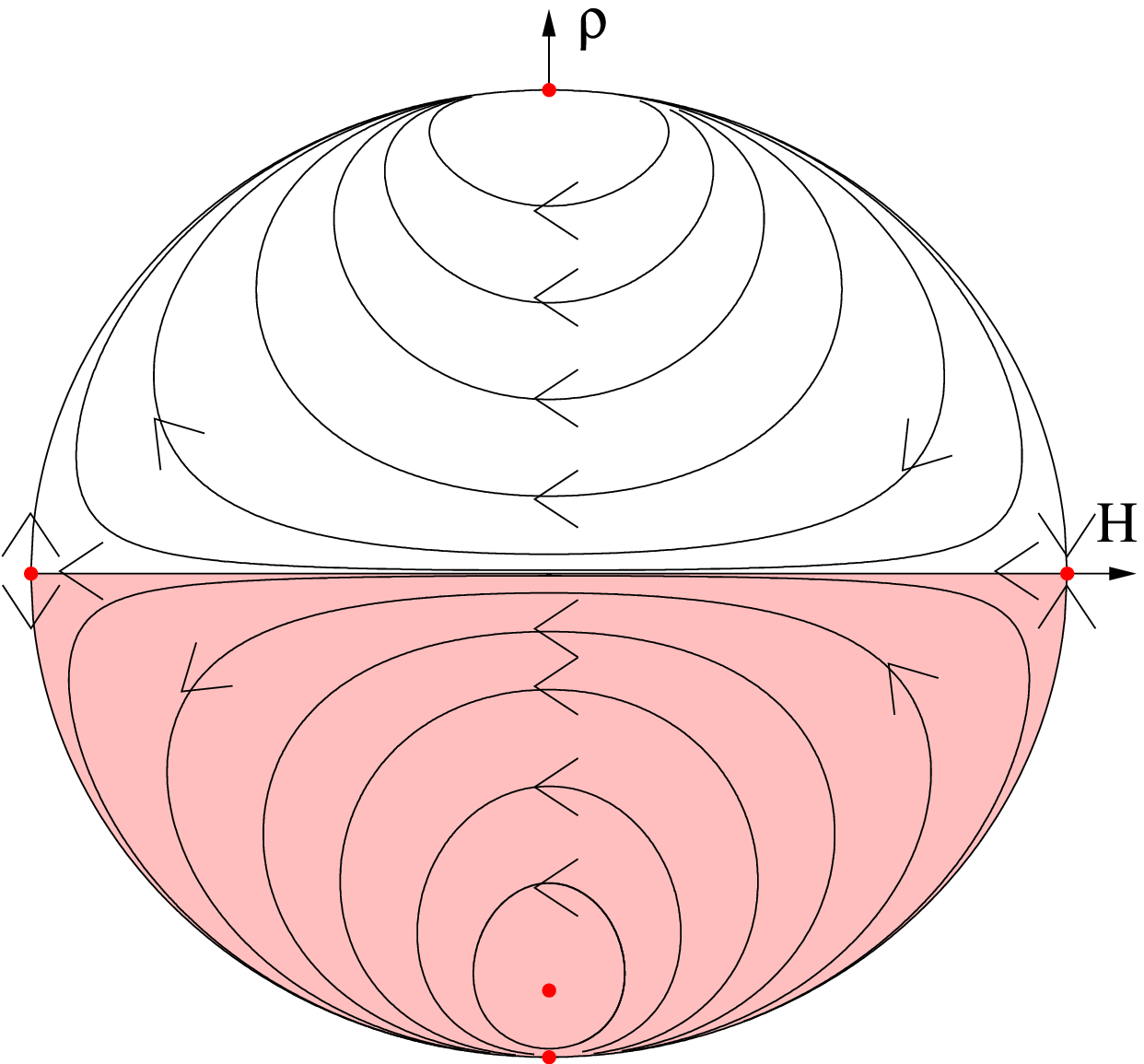} \\ [0.4cm]
\multicolumn{1}{l}{\mbox{\bf c)}} &  
\multicolumn{1}{l}{\mbox{\bf d)}} \\ [-0.5cm]
\includegraphics[scale=0.5]{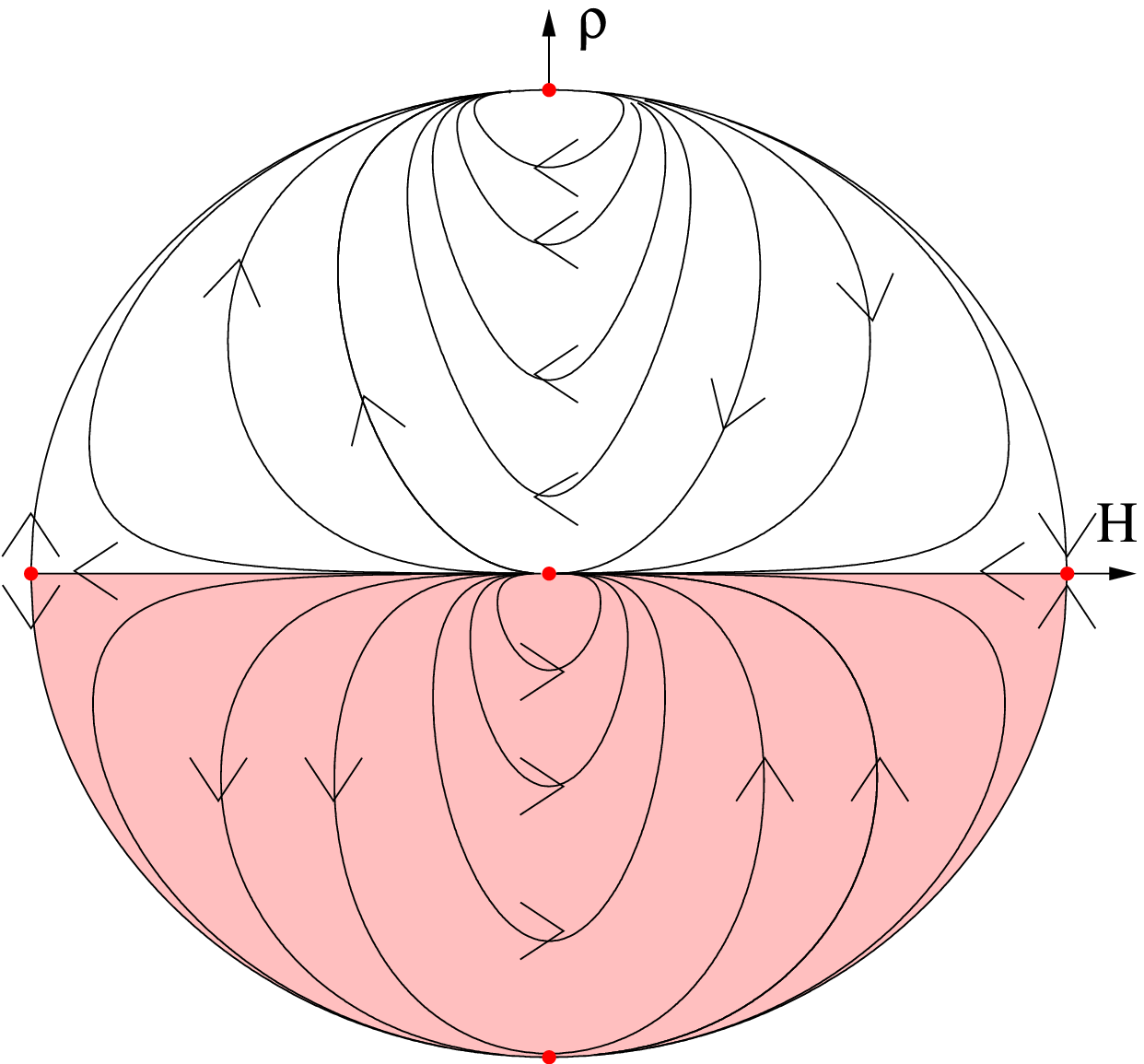} & 
\includegraphics[scale=0.5]{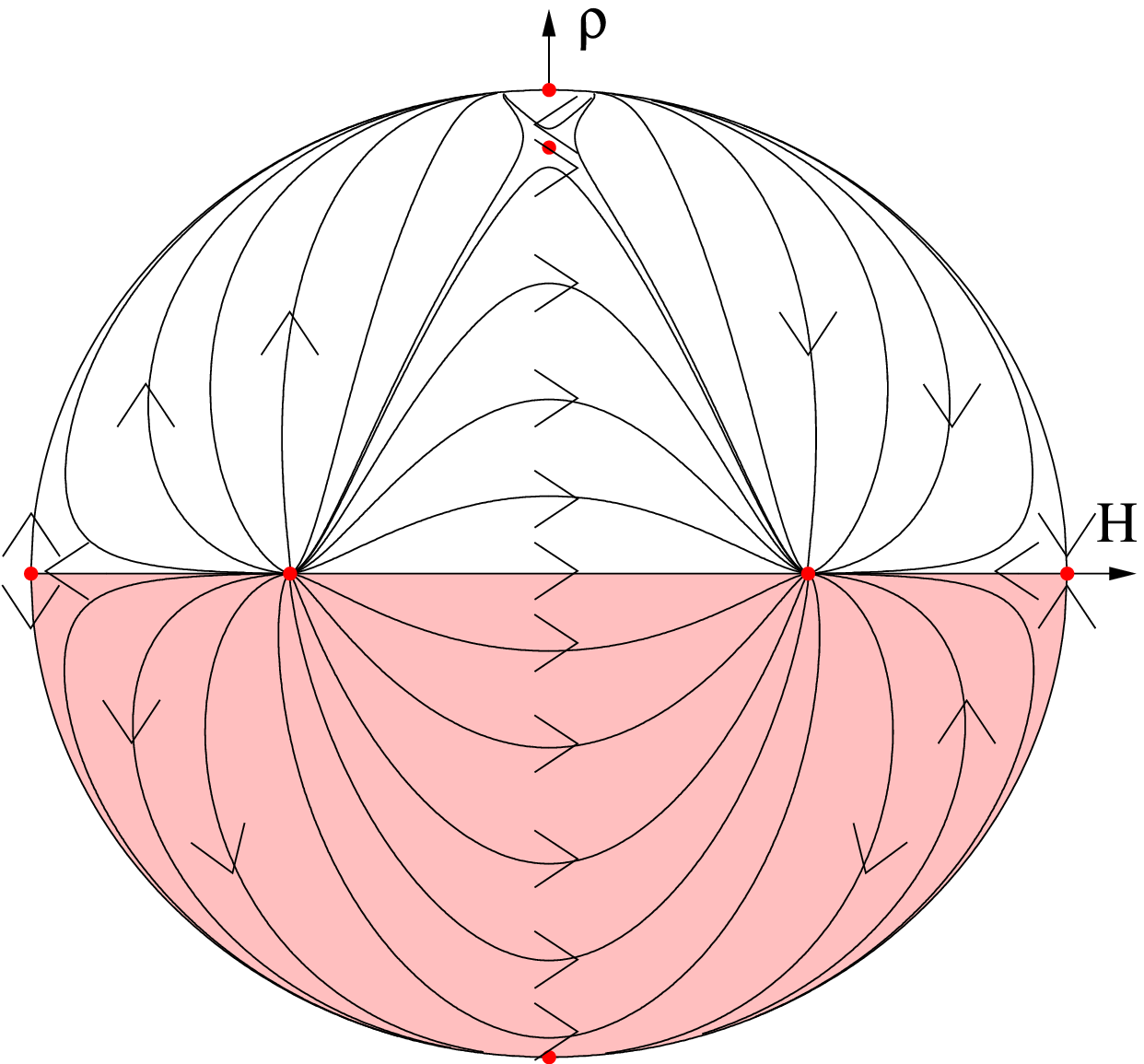} \\ [0.4cm]
\end{array}$
\end{center}
\caption{Analysis of behaviour of trajectories in phase plane transformed on the compact Poincare sphere
for {\bf a)} the FRW model with the Chaplygin gas ($A=1$,$\alpha=1$) and the dust filled FRW models with the
cosmological constant {\bf b)} $\Lambda<0$, {\bf c)} $\Lambda=0$ and {\bf d)} $\Lambda>0$.}
\label{figababbbcp}
\end{figure}
\begin{figure}[!ht]
\begin{center}
\includegraphics[scale=0.6]{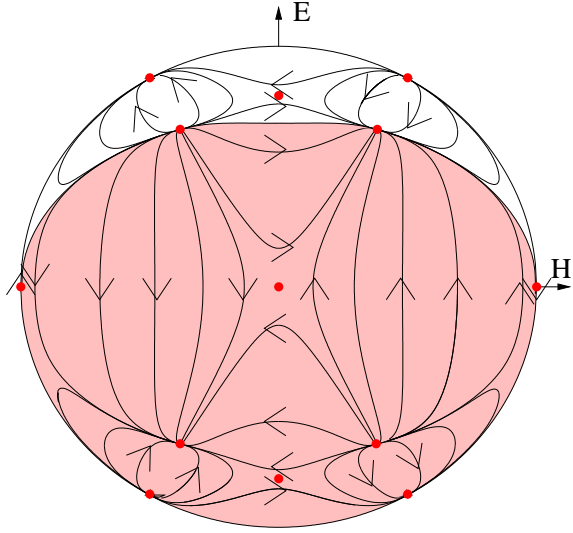}
\caption{Analysis of behaviour of trajectories for the system (\ref{newdifsys}) in phase plane $(H,E)$.
The transformation $\rho \rightarrow E = \sqrt{\rho}$ splits the composite critical point $(H=0,\rho=+\infty)$
into two simple non-degenerate points.}
\label{finp}
\end{center}
\end{figure}

The detailed analysis of the system under consideration can be performed on the phase portrait 
(Fig.~\ref{figababbbc}a). The phase trajectory for the flat model ($k=0$) is given by a parabole, displayed
symmetrically with respect to the $\rho$-axis, whose vertex changes its position depending on the value
of $A$. The behaviour of trajectories at infinity can be investigated by using the mapping of the phase plane
onto the ``Poincare sphere" which transforms the original phase plane into the projective (compact) plane.
If we consider the FRW model with the positive cosmological constant and dust then there are three critical 
points at infinity, in the physical region of the phase plane (i.e. $\rho>0$). These are two saddles
($H = \pm \infty$, $\rho = 0$) and the degenerate point ($H = 0$, $\rho = +\infty$). The transformation 
$E = \sqrt{\rho}$ splits the composite critical point into two simple non-degenerate points.

In the case of dynamical system (\ref{newdifsys}) with the right-hand sides in the form of polynomials 
of degree $n$ (maximal), one uses the projective coordinates. In this coordinates, points which at infinity 
correspond to a circle $S^{1}$ are covered by two lines $z = 0$, $-\infty < u < +\infty$, and 
$w = 0$, $-\infty < v < +\infty$.
The dynamical system in the projective coordinates $(z,u)$ and after the reparametrization of time
$d\sigma/d\tau = H^{n-1}$, assumes the form
\begin{align}
\dot{z}&\equiv \frac{dz}{d\sigma} = -z^{n+1}P\bigg(\frac{1}{z},\frac{u}{z}\bigg),\\
\dot{u}&\equiv \frac{du}{d\sigma} = z^{n}Q\bigg(\frac{1}{z},\frac{u}{z}\bigg) - uz^{n}P\bigg(\frac{1}{z},\frac{u}{z}\bigg),
\end{align}
where $n=2$ in the considered case.

The analysis of this system is carried out in standard way \cite{bogoyavlensky}. The results of such an
analysis, for dynamical system (\ref{newdifsys}) are presented on  Fig.~\ref{figababbbcp}a and Fig.~\ref{finp}.

\section{Structural stability of the FRW model with the generalized Chaplygin gas}

The idea of structural stability originated with Andronov and Pontryagin \cite{andronow}. Let us consider
the autonomous dynamical system $S'$ (i.e. its r.h.s) which is a small perturbation of original one $S$. 
The dynamical system $S'$ is said to be structurally stable if dynamical systems in the space of all 
dynamical systems it is close, in the metric sense, to $S$ or it is topologically equivalent to $S$. 

In the qualitative theory of dynamical systems instead of finding and analyzing an individual solution of a 
model, a space of all possible solutions is investigated. A property is believed to be realistic if it can 
be attributed to large subsets of models within a space of all possible solutions or if it possesses a 
certain stability, i.e., if it is shared by a slightly perturbed model. There is a wide
opinion among specialists that realistic models should be structurally stable, or even stronger, that 
everything should possess a kind of structural stability. What does the structural stability mean
in physics? The problem is in principle open in more than 2-dimensional case where according to Smale
there are large subsets generic of structurally unstable systems in the space of all dynamical systems 
\cite{smale}.
For 2-dimensional dynamical systems as in the considered case the  Peixoto theorem states that
structurally stable dynamical systems on compact manifolds form open and dense subsets in the space of all 
dynamical systems on the plane. Therefore, it is reasonable to require the model of a real 2-dimensional
problem to be structurally stable. When we consider the dynamics of 2-dimensional models then there is a 
simple test of structural stability, namely if the right-hand sides of the dynamical systems are in 
polynomial form the global phase portraits is structurally stable $S^{2}$ (${\mathbb R}$ with a Poincare 
sphere) if and only if 1) a number of critical points and limit cycles is finite, 2) each is hyperbolic and 
there are no trajectories connecting saddle points \cite{perko}. In the considered case the points at 
infinity are revealed on the projective plane. Two projective maps $(z,u)$, $(v,w)$ cover a circle at 
infinity given by $\{z = 0, -\infty < u < +\infty\}$ and $\{w = 0, -\infty < v < +\infty\}$.

The structural stability is sometimes considered as a precondition of the ``real existence". To have many 
drastically different mathematical models all of them equally well agreeing with the observational data
(taking into account final measurement errors) seems to be fatal for the empirical method of modern science
\cite{thom}. Therefore, any 2-dimensional structurally unstable models are probably not physically 
meaningful. These tendencies in contemporary cosmology inspired G. F. R. Ellis to formulate what is called 
by him the probability principle, which states that the universe model should be one that is a probable model
within the set of all universe models and the stability assumption, closely related to the above principle,
stating that the universe should be stable to perturbations.

From the analysis of phase portraits at a finite domain as well as at infinity one can conclude that model 
under consideration is structurally stable in its physical domain. Because there are no separatrices 
connecting saddle points, all critical points are hyperbolic and all critical points are non-degenerated 
and their number is finite. It should be treated as an adequate model of the real universe. Moreover, it is 
true for the more general class of models for which $\rho + p \geqslant 0$ only for 
$\rho > \rho_{\rm min} > 0$.

Let us consider now the general class of such models. Then on the strength of our assumption that the phase 
space does not contain trajectories along which the Lorentz invariability condition 
$\rho + p(\rho) \geqslant 0$ is broken in a finite time, the phase portrait can be simply obtained.

The universe evolution is described by one of the trajectories lying in the domain $\rho > \rho_{\rm min}$
and there is no trajectories which intersect $\partial S$ in a finite value of time. Then Universe accelerates
because the strong energy condition is violated. Other energy conditions may be satisfied at all times.

\begin{figure}[!ht]
\begin{center}
\includegraphics[scale=0.8]{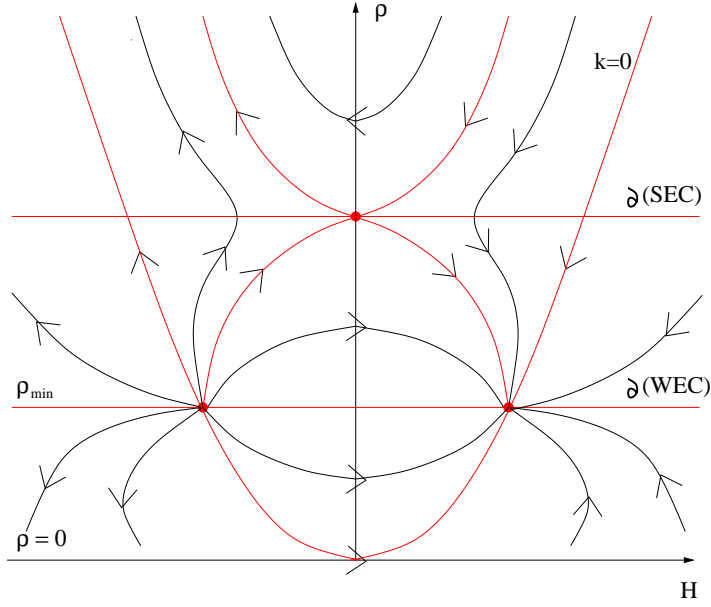}
\caption{The phase portrait for the general class of FRW models with condition that $\rho + p \geqslant 0$
for $\rho > \rho_{\rm min} > 0$. The generalized Chaplygin gas can be treated as a special case of such 
fluids. The special form of equation of state is not required because the energy condition determines the 
critical points as well as its character.
}
\label{figgen}
\end{center}
\end{figure}

Let us note that structure of the phase space does not depend on the present form of pressure because the 
critical points and their stability depend on the energy condition. Of course, the obtained phase portrait
is equivalent to that obtained for the Chaplygin gas model.

\section{FRW models with the Chaplygin gas as a Hamiltonian systems}

It is well known that the first integral FRW equation $\rho - 3H^{2} = 3k/a^{2}$ can be used to construct
a Hamiltonian function for the system. We take advantage of this feature in the considered model.
The right-hand side of the Raychaudhuri equation (\ref{difsys1}) can be expressed in terms of the scale 
factor $a(t)$ as
\begin{equation}
\ddot{a} = \frac{B}{6}a^{-2}\Bigg(2\frac{A}{B}a^{3(1+\alpha)}-1\Bigg)\Big(Aa^{3(1+\alpha)}+B\Big)^{-\frac{\alpha}{1+\alpha}}.
\label{rayeqn}
\end{equation}
Therefore the universe is accelerating provided that 
$a~>~a_{\rm cr} = \big(\frac{B}{2A}\big)^{\frac{1}{3(1+\alpha)}}$.

Equation (\ref{rayeqn}) can be rewritten in the form analogous to the Newton equation of motion with
the 1-dimensional configuration space $\{a \colon a \in {\mathbb R}_{+}\}$
\begin{equation}
\ddot{a} = -\frac{\partial V(a)}{\partial a},
\label{eqnewton}
\end{equation}
where the potential function
\begin{equation}
V(a) = -\frac{1}{6}\rho a^{2} + V_{0},
\label{eqnewtonpot}
\end{equation}
where $\rho(a)$ is given by (\ref{chapqa}); in any case $\rho$ corresponds to the effective energy density
and $V_{0} = {\rm const.}$

The first integral of (\ref{eqnewton}) is
\begin{equation}
V(a) + \frac{\dot{a}^{2}}{2} = V_{0} - \frac{k}{2}.
\label{eqnewtfint}
\end{equation}
Now, we construct the Hamiltonian function
\begin{equation}
{\mathcal H} \equiv \frac{\dot{a}^{2}}{2} + V(a).
\label{hamiltonian}
\end{equation}
Therefore the trajectories of the system lie on the energy level ${\mathcal H} = E = {\rm const.}$

It is useful to consider dynamics on the zero-energy level ${\mathcal H} = E = 0$. Then we have 
$V_{0} = k/2$ and the Hamiltonian constraint coincides with the form of the first integral.

Finally we obtain the potential of the system in the following form
\begin{equation}
V(a) = -\frac{1}{6}\Bigg(A+\frac{B}{a^{3(1+\alpha)}}\Bigg)^{\frac{1}{1+\alpha}}a^{2}+\frac{k}{2}
 = -\frac{\rho_{\rm eff}}{6}a^{2}.
\label{potfunc}
\end{equation}
There are many advantages of representing dynamics as a 1-dimensional Hamiltonian system, namely
\begin{enumerate}
\item
we can discuss the existence and stability of critical points only in terms of geometry of the potential 
function;
\item
the presented formalism gives us a natural base to discuss (in terms of the potential function) how 
cosmological problems like the horizon problem or flatness problem are solved by models;
\item
due to the existence of Hamiltonian constraint the classification of all possible evolutional paths can be 
performed by the consideration of the equation for the boundary of domain admissible for motion
\begin{equation*}
{\mathcal D} = \{a \colon V(a) \leqslant 0\}.
\end{equation*}
\end{enumerate}
Let us comment now some of advantages.

In our case the Hamiltonian function takes the simplest form for natural mechanical systems (i.e. the kinetic
energy is quadratic in velocities and the potential function depends only on generalized coordinates). Then,
the possible critical points in a finite domain of phase space are only centres or saddles. In this case
the characteristic equation is $\lambda^{2} + \partial^{2}V/\partial a^{2}|_{0} = 0$ (at the critical point
$a = a_{0}$ we have $\partial V/\partial a|_{a=a_{0}} = 0$ and $\dot{a}|_{a=a_{0}} = 0$). Therefore if 
$\partial^{2}V/\partial a^{2}|_{0} > 0$ then critical points are centres, i.e. a model is structurally 
unstable. One can easily check that in our case $V''(a) < 0$ which gives that only saddles points 
are admitted and the model is structurally stable.

The horizon problem is solved when $\dot{a} \leqslant {\rm const.}$ (may be equal to zero) as we go to 
singularity $a \rightarrow 0$. Then the distance to the horizon $\int a^{-1}dt$ diverges which means that 
there are no causally disconnected regions as $a \rightarrow 0$. Now, from the Hamiltonian constraint we 
obtain that $V(0) = 0$ or in the vicinity of singularity $V > -C^{2}/2$. Therefore the divergence of the 
potential function near the singularity indicates the presence of the particle horizon in the past.

From the physical point of view it is interesting along which trajectories the acceleration condition
$\ddot{a} > 0$ is satisfied. One can easily observe this phenomenon from the geometry of the potential 
function. If the potential function is a decreasing function of the scale factor then the universe 
accelerates. Therefore the accelerating region is situated to the right from the saddle point or for the 
Lemaitre-Einstein universes the acceleration begins in the middle of quasistatic stage. In order the solution
of the flatness problem means that at large $a$ the universe accelerates, such that we can find constant $C$ 
that $\dot{a} \geqslant C$ (then $\ddot{a} > 0$ at large $a$). Hence $V < -C^{2}/2$ at large $a$ from the 
Hamiltonian constraint.

Let us note that near the initial singularity $V(a) \propto 1/a$ like for the dust matter and $V$ goes to 
minus infinity as we are going to the initial singularity. Hence, the horizon problem is not solved by the 
model. On the other hand for large $a$, $V(a) \propto a^{2}$ and the flatness problem is solved. Let us now 
concentrate on the possibility of classification of evolutional paths by analyzing the characteristic curve 
which represents the boundary equation of the domain admissible for motion of the system in the configuration
space. Because $\dot{a}^{2} = -2V(a)$ the trajectories of the system lie in the region 
${\mathcal D} = \{a \colon V(a) \leqslant 0\}$. The boundary of this domain is 
$\partial {\mathcal D} = \{a \colon V(a)=0\}$.

From (\ref{potfunc}) the constant $A$ can be expressed as a function of $a$ as
\begin{equation}
A(a) = \Bigg(\frac{3k}{a^{2}}\Bigg)^{1+\alpha} - \frac{B}{a^{3(1+\alpha)}}.
\label{bounda}
\end{equation}

\begin{figure}[ht]
\begin{center}
$\begin{array}{c@{\hspace{0.2in}}c}
\multicolumn{1}{l}{\mbox{\bf a)}} & 
\multicolumn{1}{l}{\mbox{\bf b)}} \\ 
\includegraphics[scale=0.5]{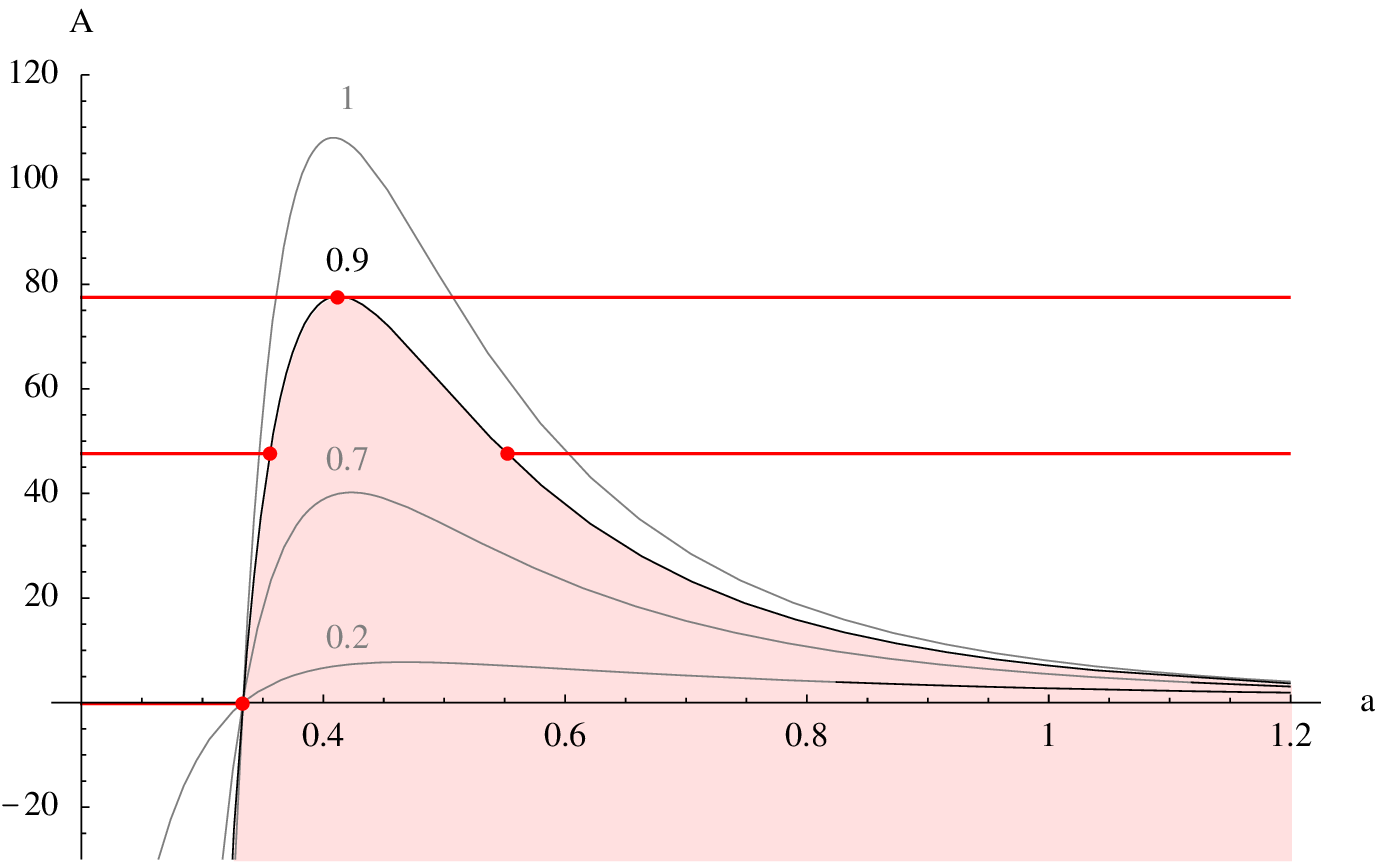} & 
\includegraphics[scale=0.5]{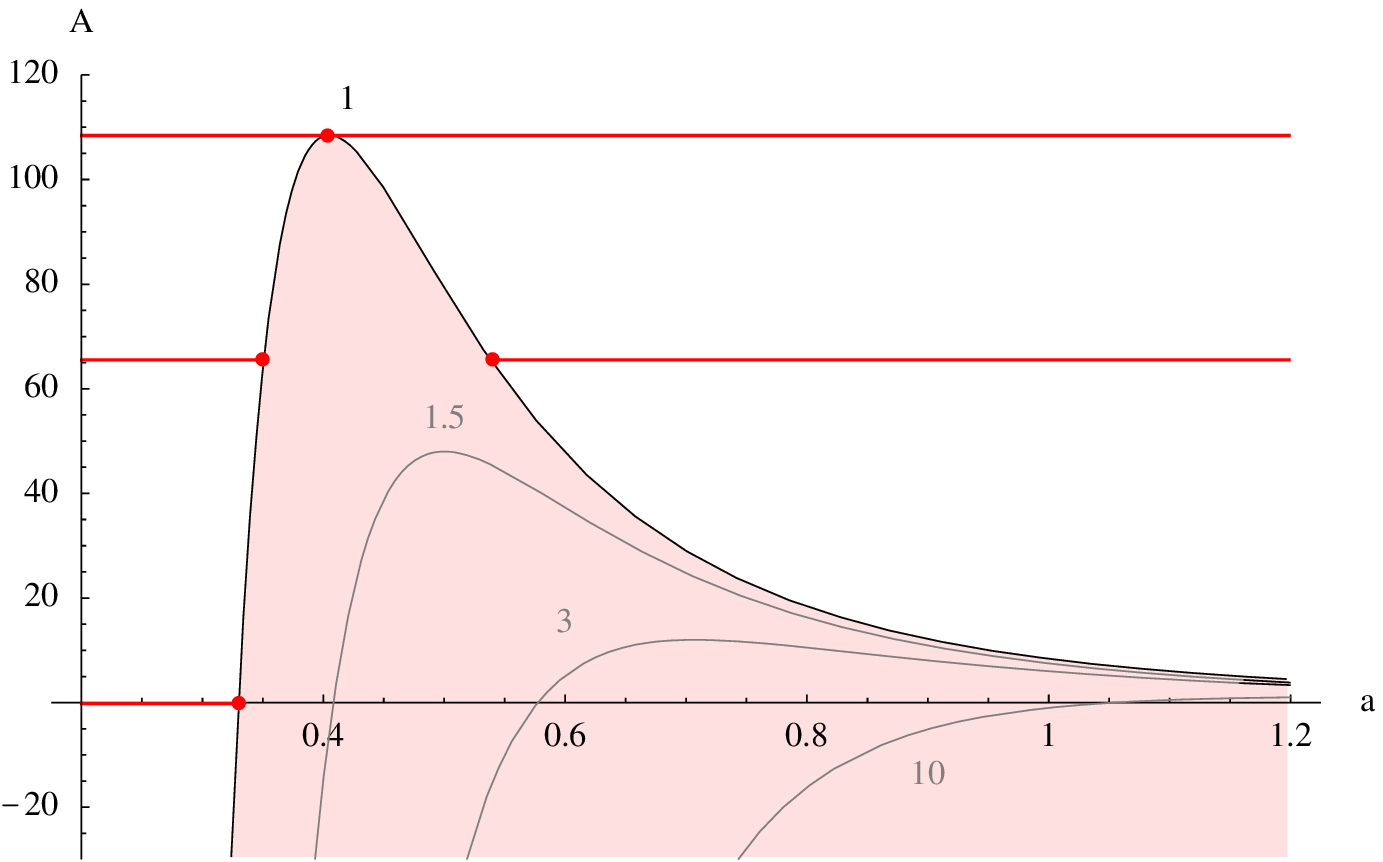} \\ [0.1cm]
\multicolumn{1}{l}{\mbox{\bf c)}} & 
\multicolumn{1}{l}{\mbox{\bf d)}} \\ 
\includegraphics[scale=0.5]{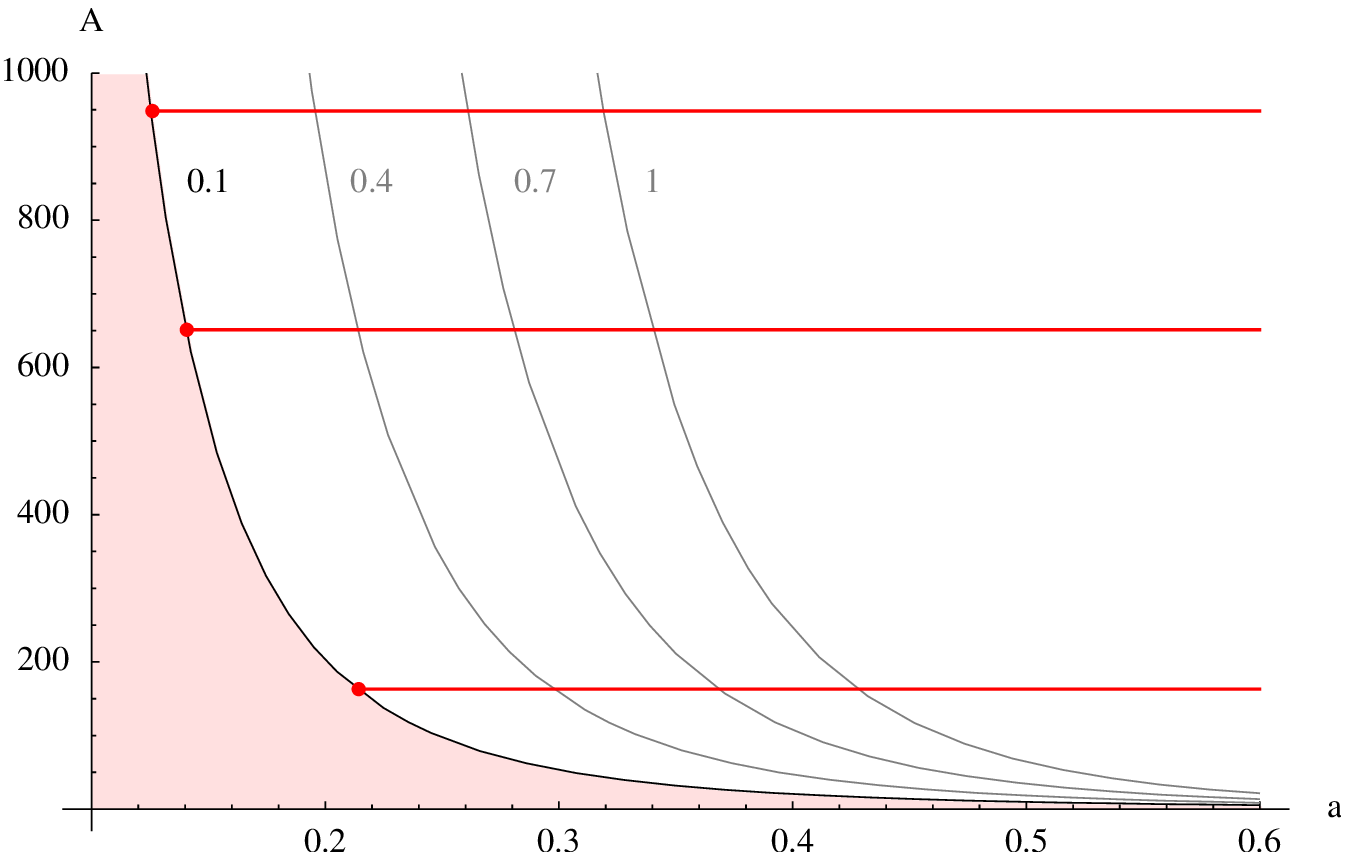} & 
\includegraphics[scale=0.5]{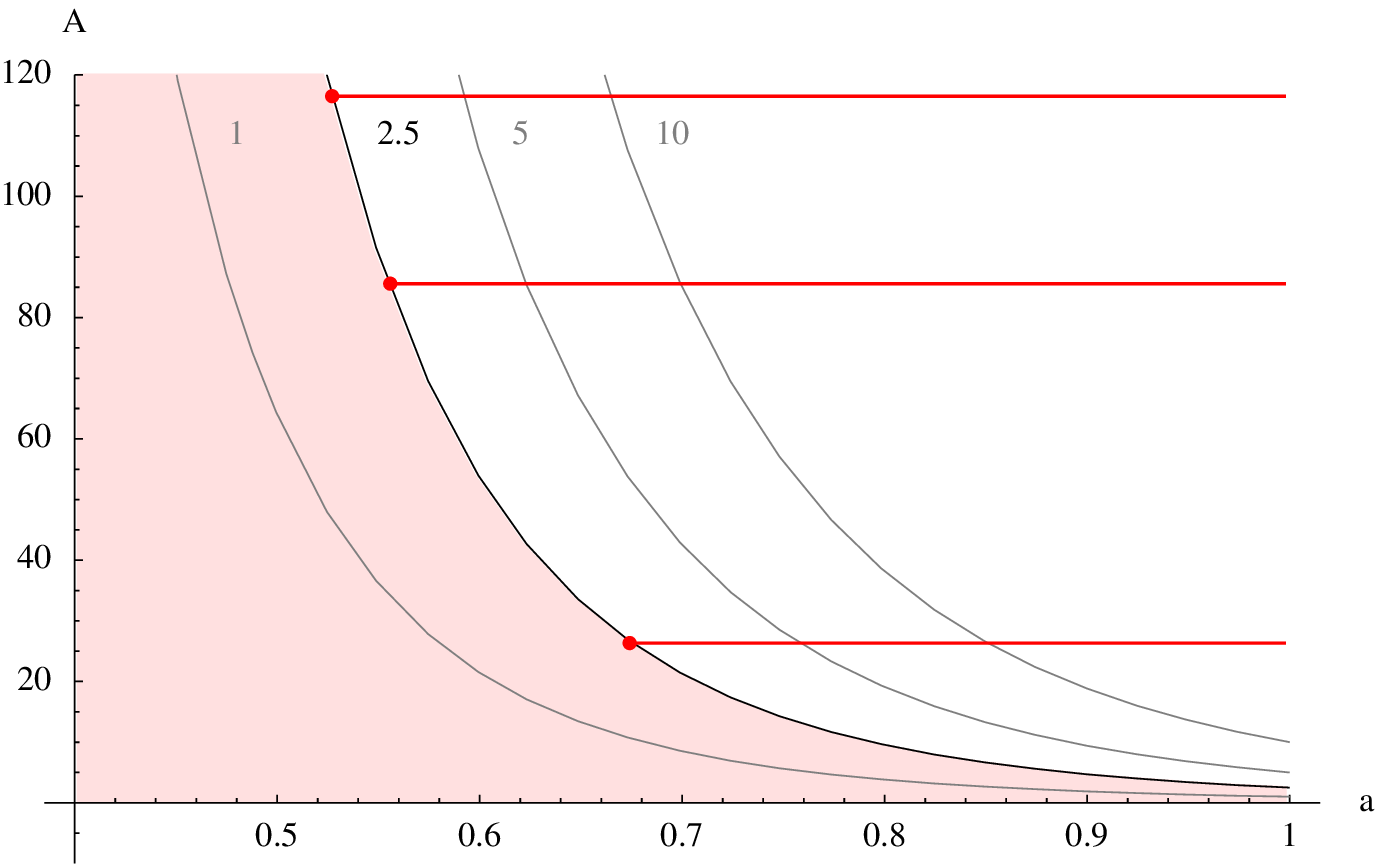} \\ [0.1cm]
\end{array}$
\end{center}
\caption{The plot of $A(a)$ (formula (\ref{bounda})) for the classification of models with potential function
(\ref{potfunc}) for {\bf a)} $B=1$, $k=+1$, $\alpha=0.2,0,7,0.9,1$, {\bf b)} $\alpha=1$, $k= \pm 1$, $B=1,1.5,3,10$, {\bf c)} $B=1$, $k=0$, $\alpha=0.1,0.4,0.7,1$ and {\bf c)} $\alpha=1$, $k=0$, $B=1,2.5,5,10$.
By considering levels of constant $A(a)$ we obtain qualitative evolution paths. The shaded region is
nonphysical because the Hamiltonian constraint.
}
\label{fig1234}
\end{figure}

The plot of $A(a)$ for different $\alpha$ and $B$ is shown in Fig.~\ref{fig1234}. 
Finally we consider the evolutional paths as levels of $A = {\rm const.} > 0$ and then we classify all 
evolutions modulo their quantitative properties. 
Let us note that Hamiltonian (\ref{hamiltonian}) can be rewritten in the new form using dimensionless
quantities
\begin{equation*}
x \equiv \frac{a}{a_{0}},\hspace{3mm} t \rightarrow T \equiv |H_{0}|t.
\end{equation*}
Then the basic dynamical equations are
\begin{gather}
\frac{\dot{x}^{2}}{2} = \frac{1}{2}\Omega_{k,0} + \frac{1}{2}\Omega_{{\rm Chapl},0}\Bigg(A_{s}+\frac{1-A_{s}}{x^{3(1+\alpha)}}\Bigg)^{\frac{1}{1+\alpha}}x^{2} = -V(x), \\
\ddot{x} = -\frac{\partial V}{\partial x}.
\label{dimlessdynsys}
\end{gather}
Of course, the above dynamical system has the Hamiltonian
\begin{equation*}
\mathcal{H} = \frac{p_{x}^{2}}{2} + V(x),
\end{equation*}
which should be considered on the zero-energy level.

As an example of application of these equations, to study the problem of cosmic acceleration, consider the 
case of $\Omega_{k,0}$ and $\Omega_{{\rm Chapl},0} \neq 0$. It emerges that at present our universe 
accelerates provided that $\ddot{x} > 0$ for $x = 1$, i.e. $A_{s}>1/3$. In general the domain of acceleration
depends only on the $A_{s}$, namely
\begin{equation*}
-(1-A_{s})x^{-3(1+\alpha)}+2A_{s}>0.
\end{equation*}
Therefore there is the minimum value of $x$ for which the universe accelerates, i.e.
\begin{equation*}
x > \Bigg(\frac{1-A_{s}}{2A_{s}}\Bigg)^{\frac{1}{3(1+\alpha)}}.
\end{equation*}
This value can be always expressed in terms of redshift $z$
\begin{equation*}
z < \Bigg(\frac{2A_{s}}{1-A_{s}}\Bigg)^{\frac{1}{3(1+\alpha)}} - 1.
\end{equation*}

\section{The redshift-magnitude relation for the model with generalized Chaplygin gas}

The cosmic distance measures like the luminosity distance sensitively depend on the spatial geometry and the
dynamics. Therefore, the luminosity depends on the present density parameters of different components
and their form of equation of state. For this reason a redshift-magnitude relation may be used to determine
the best parameters for the FRW model with generalized Chaplygin gas.

Let us consider an observer located at $r = 0$ at the moment $t = t_{0}$ which receives a light ray emitted
at $t = t_{1}$ from a source of the absolute luminosity $L$ located at the radial distance $r_{1}$. 
The redshift $z$ of the source is related to the scale factor at the two moments of evolution by 
$1 + z = a(t_{0})/a(t_{1}) = a_{0}/a$. If the apparent luminosity of the source as measured by the observer
is $l$ then the luminosity distance $d_{L}$ of the source is defined by the relation
\begin{equation}
l = \frac{L}{4 \pi d_{L}^{2}}
\end{equation}
where $d_{L} = (1+z) a_{0} r_{1}$.
The luminosity distance in the flat FRW model with generalized Chaplygin gas takes the form
\begin{equation}
d_{L}(z) = \frac{1+z}{H_{0}}\int_{0}^{z} \frac{dz'}{\sqrt{\Omega_{m,0}(1+z')^{3}
+\Omega_{{\rm Chapl},0}(A_{s}+(1-A_{s})(1+z)^{3(1+\alpha)})^{1/(1+\alpha)}}}.
\end{equation}
From above expression we obtain the relation between the apparent magnitude $m$ and absolute magnitude
$M$ in the form
\begin{equation}
m - \mathcal{M} = 5 \log \Bigg[(1+z)\int_{0}^{z} \frac{dz'}{\sqrt{\Omega_{m,0}(1+z')^{3}
+\Omega_{{\rm Chapl},0}(A_{s}+(1-A_{s})(1+z)^{3(1+\alpha)})^{1/(1+\alpha)}}}\Bigg],
\end{equation}
where
\begin{equation}
\mathcal{M} = M - 5 \log{H_{0}} + 25.
\end{equation}
In order to compare with the supernove data, we compute the distance modulus
\begin{equation}
\mu_{0} = 5 \log(d_{L}) + 25,
\end{equation}
where $d_{L}$ is in Mps. We define effective rest-frame B-magnitude $m_{b}^{\rm eff}$ for 54 supernovae
which relate to the HzST results through $m_{b}^{\rm eff} = M_{b} + \mu_{0}$, where $M_{b}$ is the peak
B-band absolute magnitude of a standard supernova.
\begin{figure}[!ht]
\begin{center}
$\begin{array}{c@{\hspace{0.2in}}c}
\multicolumn{1}{l}{\mbox{\bf a)}} & 
\multicolumn{1}{l}{\mbox{\bf b)}} \\ 
\includegraphics[scale=0.5]{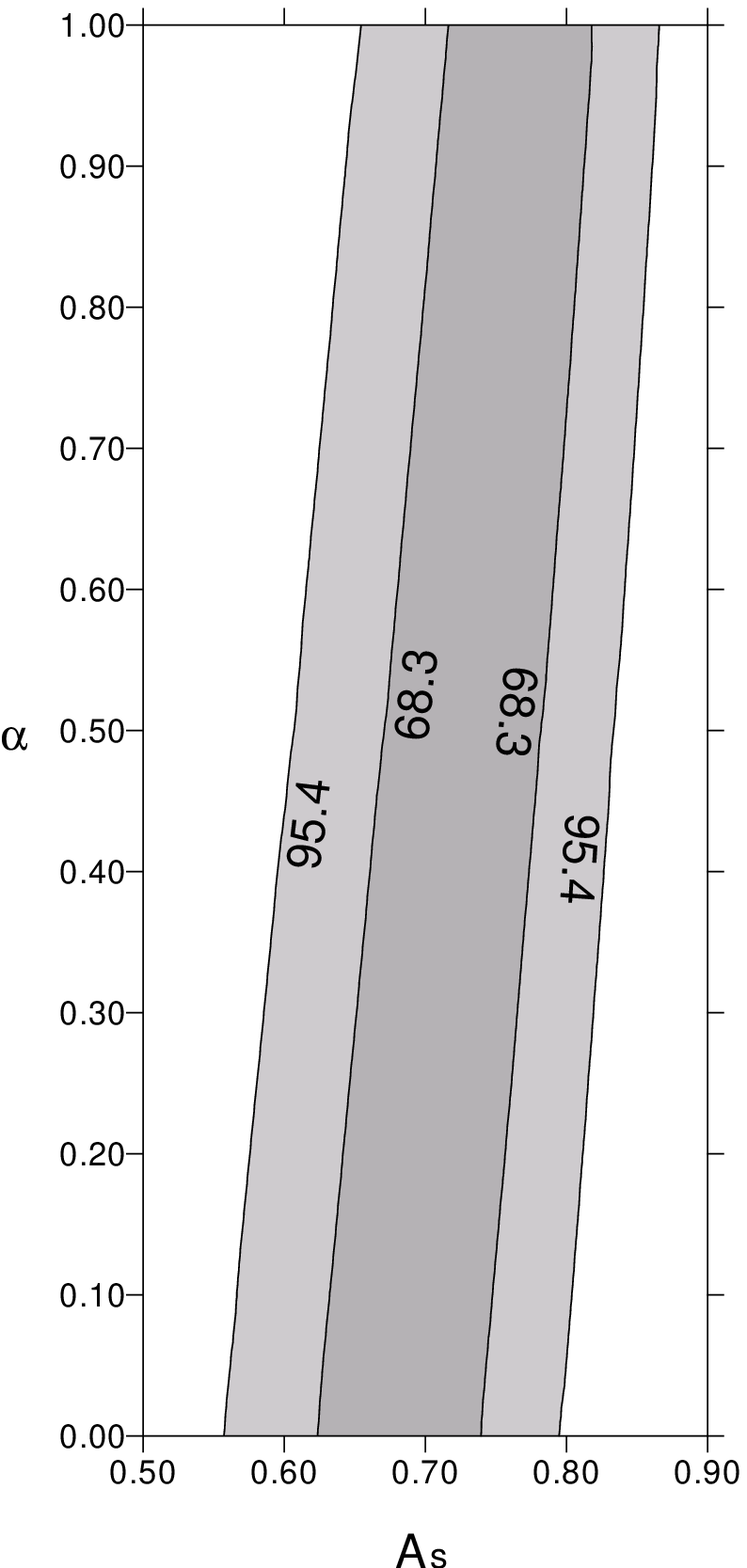} & 
\includegraphics[scale=0.5]{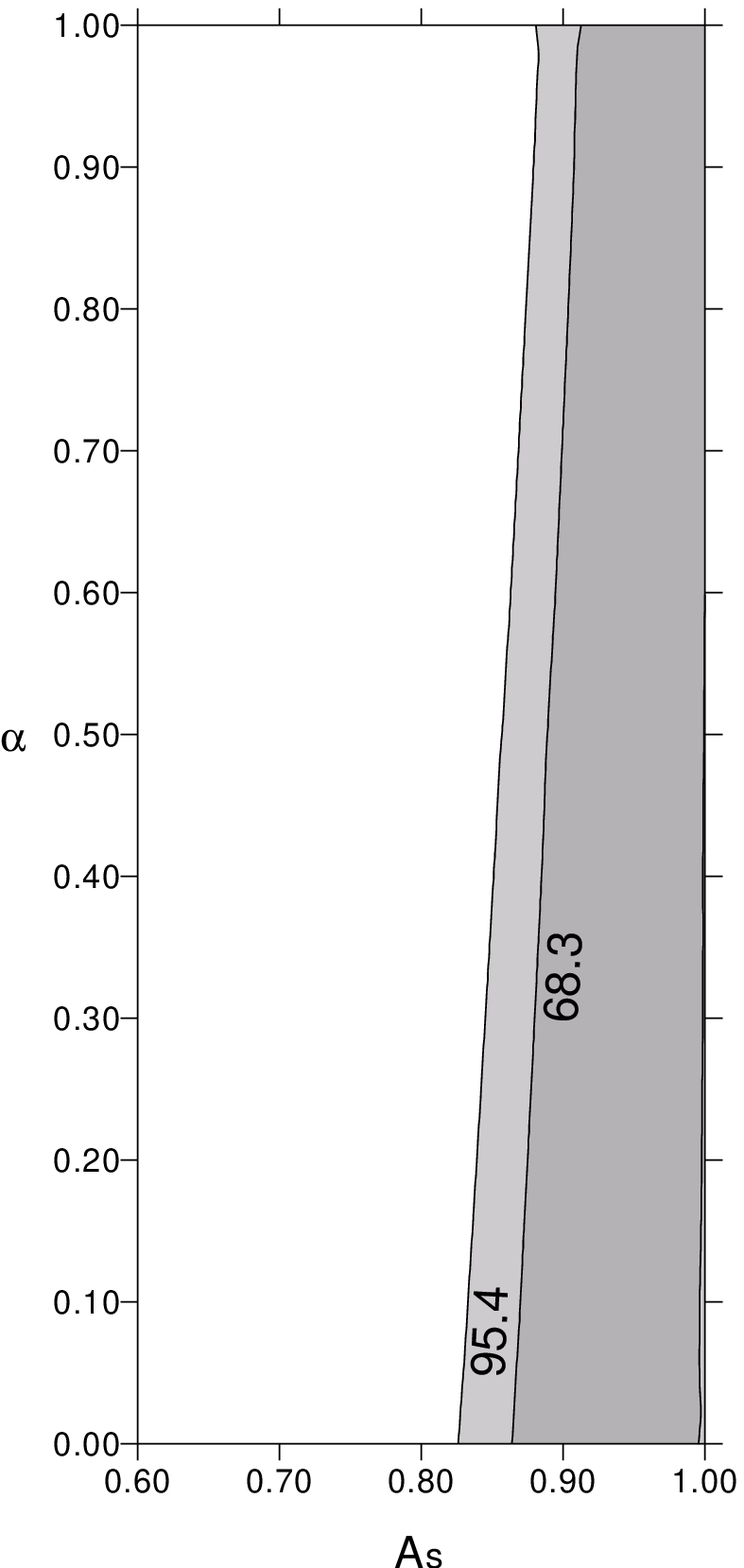} \\ [0.1cm]
\end{array}$
\end{center}
\caption{Confidence levels on the plane $(\alpha,A_{s})$ for {\bf a)} $\Omega_{m,0}=0.05$ and
{\bf b)} $\Omega_{m,0}=0.3$, ($\Omega_{m,0}+\Omega_{{\rm Chapl},0}=1$). The figures show the areas
of the preferred value of $\alpha$ and $A_{s}$. The shaded areas show the parameter regions with
confidence level $68.3\%$ ($95.4\%$).
}
\label{figconfidence}
\end{figure}
\begin{figure}[!ht]
\begin{center}
\includegraphics[scale=0.6]{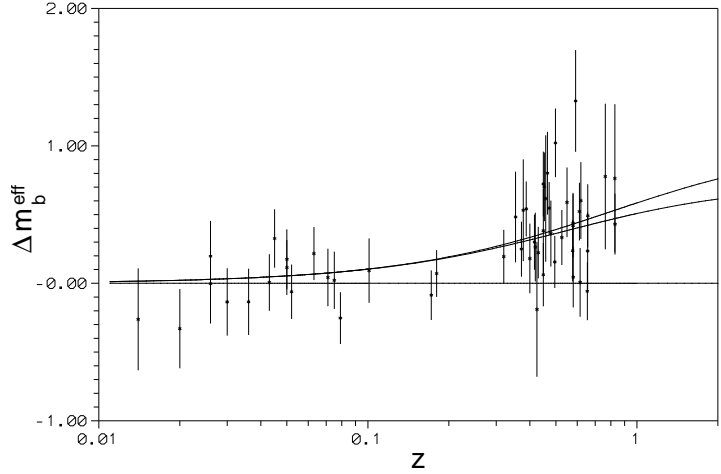}
\caption{Relative magnitude (residuals) with respect to the Einstein-de Sitter model for two cases:
the Perlmutter flat model (highest curve), and the flat model with generalized Chaplygin gas
(middle curve) for $\Omega_{m,0} = 0.05$, $\Omega_{{\rm Chapl},0} = 0.95$, $\alpha = 0.51$, $A_{s} = 0.73$,
$\chi^{2}=53.66$ (best fit parameters). For $\Omega_{m,0} = 0.3$ we obtain $\alpha = 0.95$, $A_{s} = 0.95$,
$\chi^{2}=53.64$. We fit the model with generalized Chaplygin gas to the sample C of
Perlmutter SNIa data. Sample A is the entire data set. Sample B excludes four outliers: SN1992bo and SN1992bp
(the most significant outliers from the average ligt-curve width) with lower redshifts and two with higher
redshifts, SN1994H and SN1997O (the largest residuals from $\chi^{2}$ fitting). Sample C further excludes two
very likely reddened supernovae, SN1996cg and SN1996cn.
}
\label{figrezc}
\end{center}
\end{figure}

The best fit parameters are obtained by minimizing the relation
\begin{equation}
\chi^{2} = \sum_{i}\frac{|\mu_{0,i}^{0}-\mu_{0,i}^{t}|}{\sigma_{\mu 0,i}^{2}+\sigma_{\mu z,i}^{2}}.
\end{equation}
In above expression, $\mu_{0,i}^{0}$ is the measured value, $\mu_{0,i}^{t}$ is the value calculated in the
model described above, $\sigma_{\mu 0,i}^{2}$ is the dispersion in the distance modulus due to peculiar
velocities of galaxies \cite{perlmutter}.
\begin{figure}[!ht]
\begin{center}
$\begin{array}{c@{\hspace{0.2in}}c}
\multicolumn{1}{l}{\mbox{\bf a)}} & 
\multicolumn{1}{l}{\mbox{\bf b)}} \\ 
\includegraphics[scale=0.5]{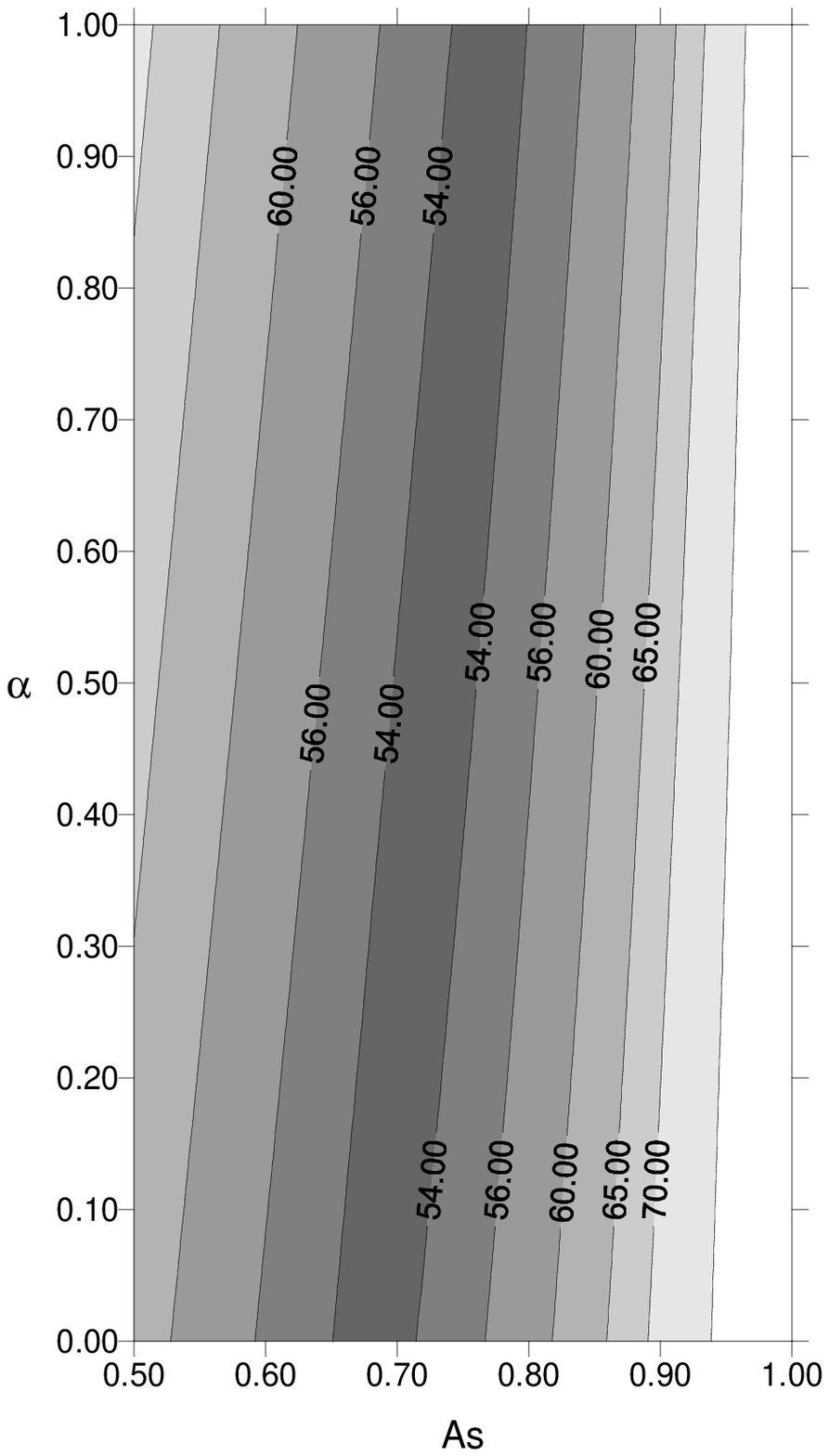} & 
\includegraphics[scale=0.5]{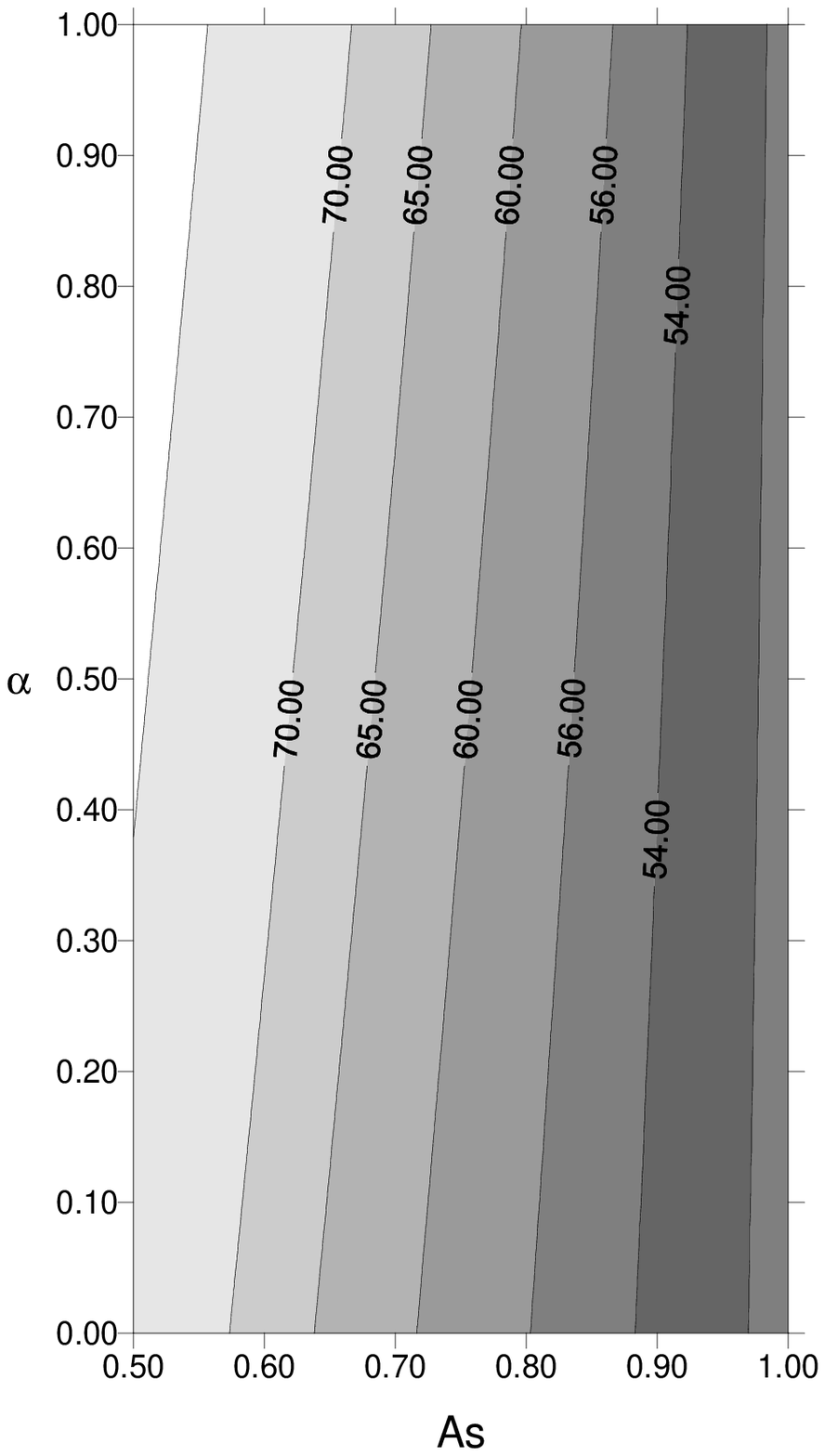} \\ [0.1cm]
\end{array}$
\end{center}
\caption{Levels of constant $\chi^{2}$ on the plane $(\alpha,A_{s})$ ($\mathcal{M}=-3.39$) with
{\bf a)} $\Omega_{m,0}=0.05$ and {\bf b)} $\Omega_{m,0}=0.3$ for the flat model ($\Omega_{k,0}=0$).
The figures show the preferred pairs $(\alpha,A_{s})$.
}
\label{fighikwadrat}
\end{figure}
We test the model with generalized Chaplygin gas using sample C of Perlmutter SNIa data. 
We may use the method of maximum likelihood parameter estimation on this data to estimate the cosmological 
parameters of interest, namely pair $(\alpha,A_{s})$. The result of statistical
analysis are presented on Fig.~\ref{figconfidence}. Figures illustrate the confidence levels as a function
of $\alpha$ and $A_{s}$ for {\bf a)} $\Omega_{m,0}=0.05$ and {\bf b)} $\Omega_{m,0}=0.3$, for the flat model 
($\Omega_{k,0}=0$).

In Fig.~\ref{figrezc} we present the plot of residuals of redshift-magnitude relationship for the supernovae 
data.

In Fig.~\ref{fighikwadrat} we show levels of constant $\chi^{2}$ on the plane $(\alpha,A_{s})$,
for {\bf a)} $\Omega_{m,0}=0.05$ and {\bf b)} $\Omega_{m,0}=0.3$,
calculated as the lowest value of $\chi^{2}$ for each pair of values $(\alpha,A_{s})$ for 
($\mathcal{M}=-3.39$). These figures show the favoured regions of pairs $(\alpha,A_{s})$.

\section{Conclusions}

In the present paper we reduce the dynamics of an universe filled with the Chaplygin gas to the form of a 
particle of unit mass in one-dimensional potential $V(a)$, where $a$ is the scale factor. Such a procedure 
gives at once an insight into the possible evolutional paths, similarly as in the classical mechanics. We 
consider the complementary description of dynamics of the model on 2-dimensional phase plane. From the 
theory of qualitative analysis of differential equations we obtain the visualization of the system evolution 
in the phase plane $(H,\rho)$, where $H$ is the Hubble function and $\rho$ -- energy density of the 
generalized Chaplygin fluid. Such a geometrization of dynamics is very useful to analyze the asymptotic 
states and their stability and to study other interesting properties of dynamical systems. Especially, it is 
interesting to study the solution of the horizon problem and initial conditions for present acceleration
of our universe. We have a neat interpretation of a domain of acceleration as a domain in configuration 
space where the potential function decreases. Therefore from the observation of the potential function,
we can find the acceleration domain $a>a_{\rm cr}$ with 
$a_{\rm cr}\colon \partial V/\partial a|_{a_{\rm cr}}=0$.
On the other hand the horizon problem is solved if $V(a)$ is bounded as $a \rightarrow 0$. If we find 
trajectories for which $V(a) \rightarrow -\infty$ as $a \rightarrow 0$ then the particle horizon in the past
is present in the model. We demonstrate that the FRW model with the Chaplygin gas does not solve the horizon 
problem whereas the flatness problem is here resolved.

There is an opinion quite widely spread among specialists that physically realistic models of the world
should be structurally stable or, even stronger, that everything that exists should possess some kind of 
structural stability \cite{smale}-\cite{andronow}. The universe certainly exists but its stability
properties are by no means clear. However In the case of 2-dimensional systems there is a simple test of 
structural stability and moreover from the Peixoto theorem such systems form open and dense subsets in the 
space of all dynamical systems on the plane, i.e. structurally stable systems are generic whereas 
structurally unstable systems are exceptional (or non-generic). If we consider FRW models with cosmological 
constant then only models with positive cosmological constant are structurally stable 
(Fig.~\ref{figababbbc}d and Fig.~\ref{figababbbcp}d).
We prove that the FRW models with the Chaplygin gas are structurally stable in their physical domains. 
This indicate that this model, from the theoretical point of view, should be seriously treated as a possible 
candidate to describe the actual universe which accelerates. We also discuss how the FRW model with the 
generalized Chaplygin gas fits the SNIa data.

\begin{acknowledgments}
The work was supported by The Rector's Scholarship Found.
\end{acknowledgments}

\end{document}